\documentclass[a4]{article} 
\usepackage{arxiv}
\usepackage[utf8]{inputenc} 
\usepackage[T1]{fontenc}    
\usepackage{booktabs}       
\usepackage{amsfonts}       
\usepackage{nicefrac}       
\usepackage{microtype}      
\usepackage{textcomp}       

\usepackage{times}
\usepackage{graphicx}
\usepackage{subcaption}                 
\usepackage{amssymb}
\usepackage{url}
\usepackage{hyperref}
\usepackage{geometry}
\usepackage{amsmath}
\usepackage{physics}
\usepackage{verbatim}                   
\usepackage[ruled,vlined]{algorithm2e}  
\usepackage{fancyhdr}
\usepackage{cleveref}   

\graphicspath{{figures/}}

\fancypagestyle{firststyle}
{
   \fancyhf{}
   \chead{}
}

\begin{document}

\title{A Hybrid Quantum-Classical CFD Methodology with Benchmark HHL Solutions}

\author{Leigh Lapworth \\
\\
Rolls-Royce plc \\
Derby, UK \\
\today
\\
\\
leigh.lapworth@rolls-royce.com  \\
}

\maketitle
\thispagestyle{firststyle}     

\begin{abstract}
There has been significant progress in the development of quantum algorithms for 
solving linear systems of equations with a growing body of applications to 
Computational Fluid Dynamics (CFD) and CFD-like problems.
This work extends previous work by developing a non-linear hybrid quantum-classical
CFD solver and using it to generate fully converged solutions.
The hybrid solver uses the SIMPLE CFD algorithm, which is common in many industrial CFD codes,
and applies it to the 2-dimensional lid driven cavity test case.
A theme of this work is the classical processing time needed to prepare the quantum circuit
with a focus on the decomposition of the CFD matrix into a linear combination
of unitaries (LCU). CFD meshes with up to $65 \times 65$ nodes are considered with the largest
producing a LCU containing 32,767 Pauli strings. A new method for rapidly re-computing
the coefficients in a LCU is proposed, although this reduces, rather than eliminates, the
classical scaling issues.
The quantum linear equation solver uses the Harrow, Hassidim, Lloyd (HHL) algorithm via
a state-vector emulator.
Test matrices are sampled from the classical CFD solver to investigate the solution
accuracy that can be achieved with HHL.
For the smallest $5 \times 5$  and $9 \times 9$ 
CFD meshes, full non-linear hybrid CFD calculations are performed.
The impacts of approximating the LCU and the varying the number of ancilla rotations
in the eigenvalue inversion circuit are studied.
Preliminary timing results indicate that the classical
computer preparation time needed for a hybrid solver is just as important to the 
achievement of quantum advantage in CFD as the time on the quantum computer.
The reported HHL solutions and LCU decompositions provide a benchmark for future 
research. The CFD test matrices used in this study are available upon request.
\end{abstract}

%
\section{Introduction}
Simulation and modelling, enabled by high performance computing, have transformed 
the way products are designed and engineered.
Successive reports and studies, e.g. \cite{walport2018computational}, have identified
computational modelling as a crucial enabler to national productivity and competitiveness.
Principle amongst these tools is Computational Fluid Dynamics (CFD) whose uses range from
designing widgets on a laptop to simulating the flow through gas turbines using classical supercomputers 
\cite{hills2007achieving, lapworth2021parallel, perez2020large, gerstenberger20213d}.

Whilst the equations governing CFD are highly nonlinear, many solvers adopt a two layer approach
where an outer non-linear step is used to linearise the equations for an inner linear
equation solver \cite{versteeg2007introduction}.
This makes CFD attractive for quantum computing particularly for large scale applications which
consume huge amounts of classical computing resources in solving the linearised equations.

Previous quantum CFD research includes \cite{steijl2019quantum} in which the Quantum Fourier
Transform was applied to the Poisson solver within a vortex in-cell method. 
The resulting hybrid methodology was simulated on a parallel computer using 8 processors 
to model colliding vortex rings on a $128^3$ mesh.
\cite{gaitan2020finding} emulated the quantum ODE algorithm of \cite{kacewicz2006almost}
to model the 1D flow through a Laval nozzle, demonstrating identical quantum and classical solutions.
\cite{costa2019quantum} modelled the 1D and 2D wave equations using emulated Hamiltonian simulations.
\cite{suau2021practical} extended the simulation of the wave equation to include the design of circuits
that were run on an ATOS QLM.
\cite{lu2020quantum} used the IBM Qiskit emulator to model the 1D heat conduction equation with 
4 mesh points using the HHL algorithm  \cite{harrow2009quantum}.
\cite{kyriienko2021solving} developed a machine learning approach based on differential
quantum circuits and simulated its application to the flow in a convergent-divergent 1D nozzle.
Of note is \cite{lubasch2020variational} where a variational algorithm was used to develop a 
NISQ era algorithm for
solving Burger's equation on the 20 qubit IBM quantum computers Tokyo and Poughkeepsie.
This is one of the few studies to solve a CFD type equation on a physical device.
\cite{ljubomir2022quantum} used the Lattice Boltzmann method (LBM) to solve the
stream-function ($\psi$) vorticity ($\omega$) formulation of the Navier-Stokes equation. 
The circuits followed the evolution of
the LBM which does not require the solution of matrix equations. 
The input state was encoded using the components of ($\psi$) and ($\omega$) along the 
lattice directions, resulting in diagonal collision and propagation operators and a simple
unitary decomposition.
The 2D lid driven was solved for a 16x16 mesh and required a total of 15 qubits  
using IBM Qiskit in emulation mode. 

More generally, significant progress has been made on Quantum Linear Equation Solvers (QLES) 
since the HHL
(Harrow-Hassidim-Lloyd) algorithm was published in 2009  \cite{harrow2009quantum}.
As in \cite{vazquez2020enhancing}, the Lie-Trotter formulae \cite{trotter1959product} or 
higher order Suzuki formulae \cite{suzuki1991general} are used to reduce the 
commutation errors in approximating the exponentiation of the matrix of interest 
as a product of exponentiated unitaries.
Qubitisation techniques 
\cite{Low2019hamiltonian, childs2012hamiltonian, kothari2014efficient, berry2015simulating, berry2018improved, babbush2018encoding} 
provide an exact implementation of a unitary decomposition
of the matrix of interest at the expense of an additional ancilla register for 
{\it preparing} the coefficients of the unitary decomposition. The prepare 
register is of size $\log_2(M)$ where $M$ is the number of unitaries in the 
linear decomposition:
\begin{equation}
  H = \sum_{j=1}^{M} \alpha_j U_j~
  \label{eqn-intro-hsum}
\end{equation}

\cite{berry2018improved} demonstrated a qubitised phase estimation implementation  
using a block-encoded walk operator as a qubiterate. 
At the end of the phase estimation the clock register contains an estimate of 
$\sin^{-1}(\frac{\lambda}{s})$ or $\pi - \sin^{-1}(\frac{\lambda}{s})$
for the eigenvalues $\lambda$, where $s$ is the sum of the unitary coefficients which
are made positive by taking any negative signs into the unitaries.
Whilst this formulation could be implemented within the HHL algorithm, a more
promising avenue is Quantum Singular Value Transformation (QSVT)
\cite{martyn2021grand, gilyen2019quantum, dong2021efficient} 
where the matrix is directly 
inverted using an operator that approximates $1/x$ and is modified when
$|x| < 1/\kappa$ to remove the singularity at $x=0$.

Common to most QLES approaches is the need to perform the decomposition in \Cref{eqn-intro-hsum}.
A widely used approach is to use 1-sparse Hermitian matrices for each $U_j$.
\cite{ahokas2004improved} gives quantum circuits for implementing 1-sparse matrices with
integer and real coefficients. 
\cite{berry2007efficient} gives an automatic algorithm for decomposing
a d-sparse matrix, $H$, into a set of 1-sparse matrices by representing $H$ as a graph
$G_H$ and colouring the edges of $G_H$ so that no two edges of a vertex share the the same 
colour. \cite{berry2007efficient} also derived $m=6d^2$ as the upper bound for the number
of 1-sparse matrices in the decomposition.
\cite{vazquez2018quantum} used a 1-dimensional colouring scheme to decompose a
tri-diagonal Toeplitz matrix into three 1-sparse matrices with non-integer coefficients.
This led to a circuit using phase and X-rotation gates with angles parameterised by the 
constants in the Toeplitz matrix.
\cite{suau2021practical} performed a direct decomposition of their 2-sparse bi-diagonal
matrix for the wave equation into two 1-sparse diagonal matrices.
Similarly, \cite{scherer2017concrete} who performed resource estimates for a 
$12,885^2$ Finite Element mesh with 332,020,680 edges, exploited the banded structure of
their FE matrix to decompose the matrix into nine banded 1-sparse matrices. 
The regular mesh and the FE discretisation
made this essentially a Toeplitz matrix with constants along each of the nine diagonals.
\cite{kothari2010efficient} and \cite{childs2010simulating} developed a {\it star}
decomposition in which each $U_j$ is a {\it galaxy} with a maximum number $m=6d$ of
galaxies. This traded a decomposition with fewer terms for one in which the $U_j$ matrices
were no longer 1-sparse.
 
A characteristic of previous work is that the decompositions involved matrices with
a small number of real valued degrees of freedom. The $8 \times 8$ Toeplitz matrix studied by
\cite{vazquez2020enhancing} needed only two real numbers to define all the non-zero entries.
In this work, more realistic matrices are considered where every non-zero entry in the matrix
is assumed to have a 
unique value. One question this work seeks to answer is, how efficiently can a CFD matrix be
decomposed into a linear combination of unitaries?
A simulated hybrid CFD solver has been developed based on the well established
SIMPLE algorithm \cite{patankar1972calculation} and using the HHL algorithm for the QLES.
Typical of many CFD solvers is the fact that while the matrix entries change during each
non-linear update, the matrix retains a fixed sparsity pattern.
A new method is presented for optimising the time taken to re-evaluate the coefficients in
the unitary decomposition each time the CFD matrix is updated.
The method uses clusters of Pauli strings that have a shared sparsity pattern
and has some resonances with the galaxy concept used by
\cite{childs2010simulating} although the methodologies used to construct the
clusters/galaxies are quite different.
The performance of the classical decomposition step is studied for the lid driven
cavity test case \cite{smith1975comparative, ghia1977study} using CFD meshes
ranging from $5 \times 5$ nodes to $65 \times 65$ nodes. 
For the two smallest meshes, complete hybrid CFD solution are performed to enable trade
studies of the effect of approximating the unitary decomposition on convergence rates. 
The number of ancilla rotations in the eigenvalue inversion circuit is also studied.
In this work, HHL returns the full state vector to the classical solver.
This allows the effect of the eigenvalue precision to be studied relative to the
exact solution. Future work will consider the effect of returning a sampled state
based on multiple observations, and on approximations in loading the input state.

This work is presented as follows.
\Cref{sec-cfd-alg} gives an overview of the current state of classical CFD algorithms and
considers routes to quantum advantage. This is supplemented by \Cref{sec-app-cfd-alg}
which gives a high level overview of the SIMPLE CFD algorithm for readers with
little prior knowledge of CFD.
\Cref{sec-test-case} describes the test case used in this study and gives key
parameters for the matrices to be solved.
\Cref{sec-hybrid-CFD} describes the emulated hybrid solver framework.
This is supplemented by \Cref{sec-app-lcu-method} which gives the method for efficient
unitary decomposition and coefficient re-evaluation.
\Cref{sec-results} presents timing results for the unitary decomposition relative to
the classical time for the CFD solver. This is relevant as the decomposition takes 
place on the classical side of the hybrid interface.
Hybrid HHL solutions are also presented and the impact of ignoring unitaries with
small coefficients is analysed.
Finally \Cref{sec-conclusion} draws conclusions and the directions for
future developments are highlighted.

%
\section{CFD algorithms}
\label{sec-cfd-alg}
\begin{figure}[h]
  \begin{center}
    \includegraphics[width=0.60\textwidth]{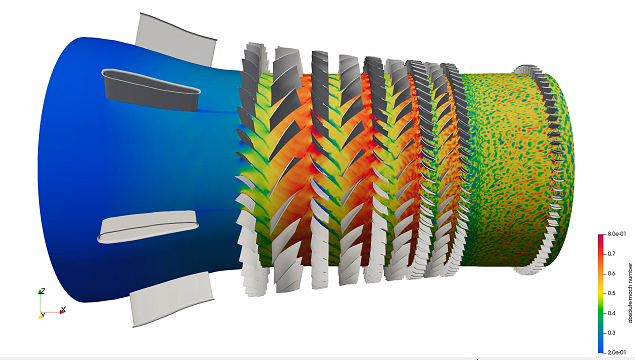}
  \end{center}
  \caption{\centering Snapshot from 5.3 billion node unsteady CFD solution from \cite{gerstenberger20213d}.}
  \label{rig250-axel01}
\end{figure}

Modern CFD codes are highly sophisticated and are able to run reliably on many millions of mesh nodes.
\cite{shahpar2020building} used steady CFD calculations with up to 180 million 
nodes to model the aerodynamic impact
of variations in the manufacture of bypass guide vanes in a gas turbine engine.
Unsteady CFD has much higher mesh requirements due to the need to resolve the relative motion
of adjacent components and features such as shocks and wakes as they propagate through 
the solution domain.
\cite{gerstenberger20213d} has demonstrated a 5.3 billion node mesh unsteady solution of
a compressor using 10,240 cores of the TU Dresden supercomputer, see \Cref{rig250-axel01}.
There are many well established techniques for the hybridisation of classical CFD codes
with many able, for example, to mix shared 
and distributed memory, and vectorization paradigms in a single code base.
\cite{reguly2020modernising} demonstrate how abstraction layers allow CFD codes to easily 
run in mixed mode and also to run across hybrid platforms such as CPUs with GPU 
accelerators. 

Whilst the power of quantum computers is expected to advance rapidly, current devices are broadly
equivalent to the classical devices of the 1970s in terms of the CFD algorithms and models they 
can process. At the heart of almost all CFD solvers is a linear equation solver. 
Although progress is being made on NISQ based CFD solvers 
\cite{kyriienko2021solving, ljubomir2022quantum}, quantum CFD solvers that can rival
the classical applications described in the previous paragraph will require fault tolerant devices. 

The flow of a fluid is governed by the Navier-Stokes equations which express
the conservation of mass, momentum and energy:

\begin{equation}
  \frac{\partial \rho}{\partial t} + \nabla \cdot (\rho \mathbf{u}) = 0
  \label{eqn-navier-rho}
\end{equation}

\begin{equation}
  \rho \frac{\partial \mathbf{u}}{\partial t} +  \nabla \cdot(\rho \mathbf{u} \otimes \mathbf{u}) = -\nabla p + \mu \nabla^{2} \mathbf{u} 
  \label{eqn-navier-u}
\end{equation}

Where $\rho$ is the fluid density, $\mathbf{u}$ is the fluid velocity, $p$ is the fluid pressure and
$\mu$ is the dynamic viscosity.
Some terms have been dropped from \Crefrange{eqn-navier-rho}{eqn-navier-u} and the energy conservation
equation has been omitted since this paper considers only incompressible, laminar flow.

To solve the Navier-Stokes equations, the SIMPLE 
(Semi-Implicit Method for Pressure Linked Equations) 
algorithm of \cite{patankar1972calculation} is adopted. This uses {\it staggered} variables on a
2-dimensional Cartesian lattice. The derivation and implementation of the 
SIMPLE algorithm are described in \Cref{sec-app-cfd-alg}.

%
\subsection{Considerations for Quantum Advantage}
\label{subsec-cfd-qa}
From the description of the SIMPLE algorithm in \Cref{sec-app-cfd-alg},
it may not be apparent that the momentum
and pressure correction equations have a different character.
The convective nature of the momentum equations means they are {\it hyperbolic}.
Whereas, the pressure correction equations are {\it elliptic}.
The latter require much more computational effort to solve.
For example, the pressure correction solution shown in \Cref{fig-4x4-bx-iter10}
required 11,849 Gauss-Seidel iterations for the pressure correction equation
compared to 33 and 36 iterations respectively for the $u$ and $v$ momentum equations.
Whilst Gauss-Seidel is a very rudimentary scheme, and modern CFD codes use far more sophisticated
solvers, these counts are indicative of the fact that the solution time in modern codes
is dominated by the pressure correction equation.
Hence the opportunities for quantum advantage are much higher for the pressure correction equation,
particularly for very large scale simulations where parallelisation across multiple nodes of a
classical supercomputer often results in a block-wise elliptic approach.

\begin{figure}[h]
  \begin{center}
    \includegraphics[width=0.50\textwidth]{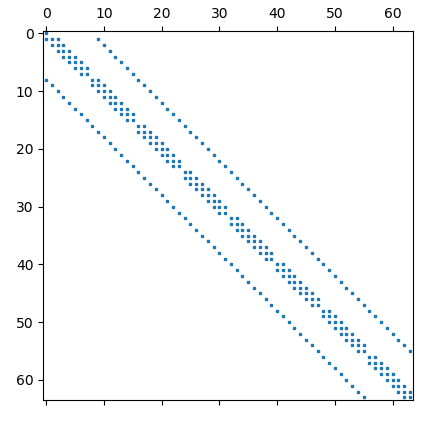}
  \end{center}
  \caption{\centering Sparsity pattern of the 8x8 pressure correction matrix for the test case
           described in \Cref{sec-test-case}.}
  \label{fig-8x8-pc-mat}
\end{figure}

Solving an inner linearised matrix or set of matrices is the standard approach used in CFD.
This gives a clear opportunity for hybrid calculations where the linearisation of the non-linear 
Navier-Stokes equations occurs on the classical computer and the resulting linear matrix system(s)
is solved on the quantum computer.
This work focuses on the hybrid interface and particularly the
decomposition of the CFD matrices into a form suitable for a quantum computer.
The approach adopted exploits the fact that there may be a large number of outer non-linear iterations
and, whilst the matrix entries are updated, the sparsity pattern does not change.
A typical sparsity pattern for the pressure correction matrix
($A^p$ in \Cref{eqn-simple-01}) on a 2D lattice mesh is shown in \Cref{fig-8x8-pc-mat}.

%
\section{Test case - lid driven cavity}
\label{sec-test-case}

The lid driven cavity is one of the canonical test cases for CFD with the first published 
applications dating from the 1970s \cite{smith1975comparative, ghia1977study}.
This case continues to be used as a basic validation case for viscous
incompressible flow solvers on coarse meshes \cite{redal2019dynamfluid}.

\begin{figure}[h]
  \begin{center}
    \includegraphics[width=0.60\textwidth]{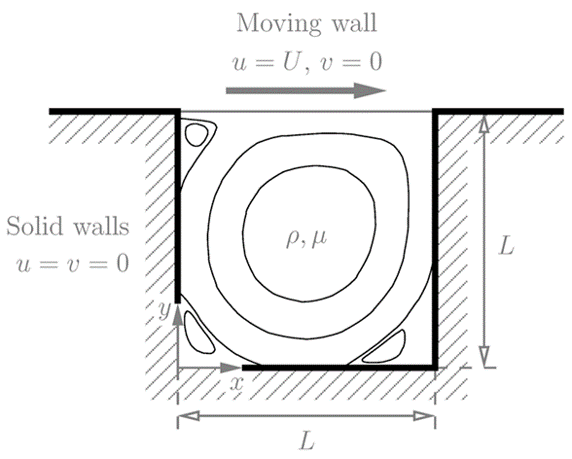}
  \end{center}
  \caption{\centering Overview of lid driven cavity test case, from \cite{redal2019dynamfluid}.}
  \label{fig-cavity-flow}
\end{figure}

Whilst, this work considers only low Reynolds number laminar flow, at higher Reynolds numbers
this test case exhibits more complicated phenomena such as recirculations,
turbulent flow structures and laminar to turbulent transition. 
\Cref{fig-cavity-flow} from \cite{redal2019dynamfluid} gives an indication of some of the
2 dimensional flow structures.
In 3D, the flow demonstrates 
complex unsteady turbulence phenomena such as inhomogeneous turbulence and small scale helical
structures \cite{bouffanais2007large, shetty2010high}.
\cite{shetty2010high} used a staggered mesh and a derivative of the SIMPLE scheme on a
$64 \times 64 \times 64$ mesh.
\cite{courbebaisse2011time} used the equivalent of a $129 \times 129 \times 129$ mesh to perform
direct numerical simulations of the turbulent flow in a 3D cavity.
Calculations of this type are attractive for quantum computing as they explicitly resolve all the
turbulence phenomena and do not require complicated physical models of turbulence.
However, this paper considers the 2D cavity with mesh sizes that are likely to be tractable on the first generation of fault tolerant devices.

\begin{table}[h]
  \centering
  \begin{tabular}{c c c c c c c c}
    \toprule
    CFD mesh & PC matrix & \#non-zeros & sparsity & $\lambda_{min}$ & $\lambda_{max}$ & $\kappa$ & HHL \\
    \midrule
     5x5     &   16x16     & 64       &  25.0\%   &$5.2\times10^{-2}$ & 4.54      &  87.7             & 15\\
     9x9     &   64x64     & 288      &  7.03\%   &$2.7\times10^{-3}$ & 1.51      & $5.7\times10^{2}$ & 19\\
     17x17   &  256x256    & 1,216    &  1.86\%   &$1.4\times10^{-4}$ & 0.49      & $3.5\times10^{3}$ & 24\\
     33x33   & 1,024x1,024 & 4,992    &  0.48\%   &$7.3\times10^{-6}$ & 0.13      & $1.8\times10^{4}$ & 29\\
     65x65   & 4,096x4,096 & 20,224   &  0.12\%   &$3.8\times10^{-7}$ & 0.034     & $8.9\times10^{4}$ & 33\\
    \bottomrule
  \end{tabular}
  \caption{\centering Dimensions, sparsity, eigenvalue range and condition number for pressure correction equations.
  The HHL column gives an estimate of the number of logical qubits needed by HHL.}
  \label{tab-pc-sparsity}
\end{table}

\Cref{tab-pc-sparsity} lists the mesh dimensions, based on the $(x,y)$ coordinate mesh, used in this
study. Also included are the dimensions of the pressure correction matrices and their sparsity as a
percentage of the number of non-zero entries relative to a fully populated matrix. 
\Cref{tab-pc-sparsity} also includes the minimum and maximum eigenvalues and the condition number,
$\kappa$ for each matrix, computed using the GNU scientific library \cite{gough2009gnu}.  
For each mesh, the eigenvalues are taken from the pressure correction equation after 10 outer
iterations. The table also gives an estimate of the number of logical qubits needed by HHL
based the the minimum and maximum eigenvalues. The time taken to compute these is not included
in the analysis.

\begin{figure}[h]
  \centering
  \begin{subfigure}[b]{0.4\textwidth}
      \centering
      \includegraphics[width=\textwidth]{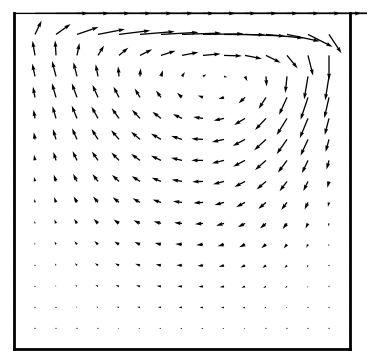}
      \caption{\small Converged velocity vectors.}
      \label{fig-17x17-vectors}
  \end{subfigure}
  \hfill
  \begin{subfigure}[b]{0.5\textwidth}
      \centering
      \includegraphics[width=\textwidth]{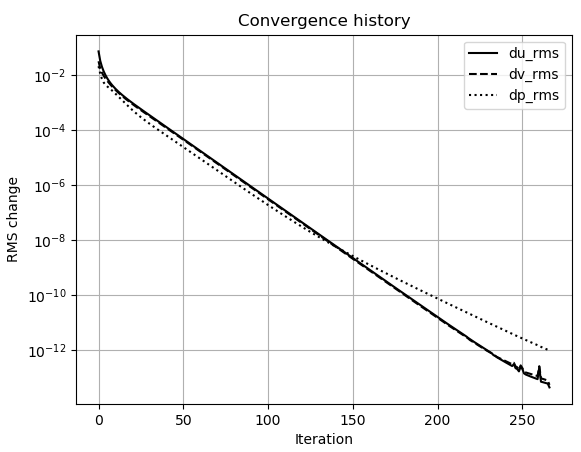}
      \caption{\small Convergence history.}
      \label{fig-17x17-hist}
  \end{subfigure}
  \caption{\centering Lid driven cavity: solution and convergence history for 17x17 mesh}
  \label{fig-cavity-soln}
\end{figure}

\Cref{fig-17x17-vectors} show the velocity vectors from the converged CFD solution
on the 17x17 mesh.
\Cref{fig-17x17-hist} shows the convergence histories for the momentum and pressure
correction equations as the root mean square (RMS) of the updates to each equation.
The calculation is performed using double precision arithmetic and the iterations are
deemed converged when the RMS updates are all below $10^{-12}$.

\begin{figure}[h]
  \centering
  \begin{subfigure}[b]{0.45\textwidth}
      \centering
      \includegraphics[width=\textwidth]{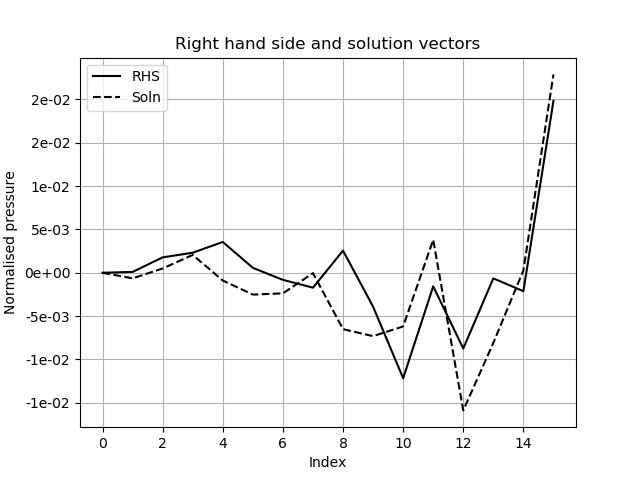}
      \caption{\small 10 iterations.}
      \label{fig-4x4-bx-iter10}
  \end{subfigure}
  \hfill
  \begin{subfigure}[b]{0.45\textwidth}
      \centering
      \includegraphics[width=\textwidth]{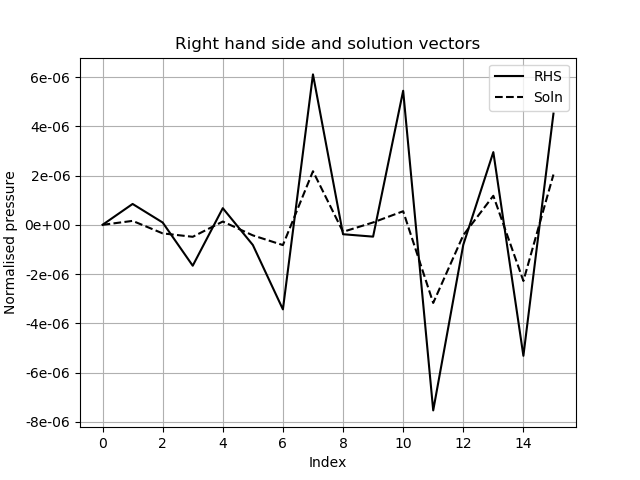}
      \caption{\small 100 iterations.}
      \label{fig-4x4-bx-iter100}
  \end{subfigure}
  \caption{\centering Lid driven cavity: right-hand side and solution vectors of the pressure
           correction equation after 10 and 100 iterations on the $5 \times 5$ mesh.}
  \label{fig-cavity-bx}
\end{figure}

The initial analysis will consider pressure correction matrices sampled from the
classical CFD solver after 10 and 100 outer iterations. HHL solutions will be
compared with the classical solutions.
\Cref{fig-4x4-bx-iter10} shows the right hand side input vector and the
solution vector for the the pressure correction matrix after 10 iterations
from the $5 \times 5$ mesh.
\Cref{fig-4x4-bx-iter100} shows the same vectors for the matrix after 100 iterations. 
Similar matrices and vectors for all the meshes listed in \Cref{tab-pc-sparsity} are
available on request.

%
\section{Hybrid CFD solver}
\label{sec-hybrid-CFD}

\begin{figure}[h]
  \begin{center}
    \includegraphics[width=0.80\textwidth]{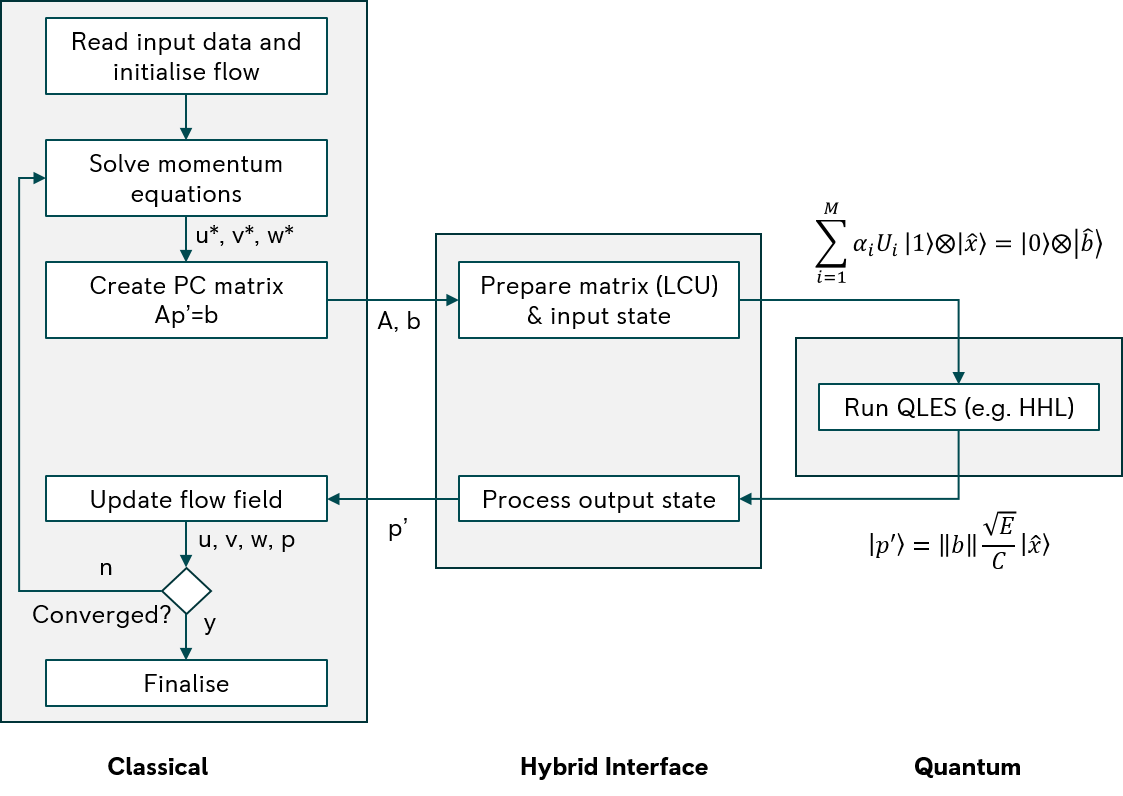}
  \end{center}
  \caption{\centering Hybrid CFD solver for the SIMPLE algorithm. Hats denote normalised variables.}
  \label{fig-hybrid-cfd}
\end{figure}

\Cref{fig-hybrid-cfd} gives an overview of the hybrid CFD solver whereby the
time consuming pressure correction equation would be solved on a quantum device.
In this work, the Quantum Linear Equation Solver (QLES) is emulated on a classical
device.

%
\subsection{Matrix preparation}
\label{subsec-matrix-prep}

In order to process the CFD matrix on a quantum device it must be decomposed into
a linear combination of unitaries (LCU). The fixed sparsity pattern of the pressure correction
matrix means that the same set of unitaries can be used throughout the hybrid
solver with only the LCU coefficients updated each time the hybrid interface is called.

The linearised system shown in \Cref{fig-hybrid-cfd} solves a matrix equation
of the form  $A\ket{x}=\ket{b}$. 
As discussed in \Cref{sec-app-lcu-method}, the matrix $A$ is non-symmetric and in order to
create a unitary decomposition it must be {\it symmetrised} such that:

\begin{equation}
  H =
  \begin{pmatrix}
    0           & A \\
    A^{\dagger} & 0
  \end{pmatrix}
  \label{eqn-A2H}
\end{equation}

And the linear system to be solved becomes:
\[
H 
  \begin{pmatrix}
    0 \\
    x
  \end{pmatrix}
  =
  \begin{pmatrix}
    b \\
    0
  \end{pmatrix}
\]

The task for the classical part of the hybrid solver is to express $H$ as a linear
combination of $M$ unitaries:

\begin{equation}
  H = \sum_{i=1}^{M} \alpha_i U_i
  \label{eqn-lcu01}
\end{equation}

In this work, a new approach based on the orthogonality of grand sums of Hadamard products 
of Pauli strings is used. This is described in full in \Cref{subsec-app-pauli-hadamard}
which also describes other LCU approaches including those
based on the trace orthogonality of products of Pauli matrices, commonly termed Pauli strings.

\Cref{tab-app-decomp-times} compares the times for the initial LCU decomposition and
for the re-evaluation of the coefficients using the trace and Hadamard orthogonality
approaches.
For validation, both approaches produce an identical number
of unitaries in each decomposition. 
For the Hadamard based approach, the numbers in brackets indicate the number of
clusters that were found.
Both decompositions are compared with the original matrix to
confirm that they are equivalent. In all cases the $L_2$ norm of the differences
between the original and the unitary decomposition are $\mathcal O(10^{-15})$.

\begin{table}[h]
  \centering
  \begin{tabular}{c c c c c c c}
    \toprule
    \multicolumn{1}{c}{Dimensions} & \multicolumn{3}{c}{Trace orthogonality} & \multicolumn{3}{c}{Hadamard orthogonality} \\
    \cmidrule(r){1-1}
    \cmidrule(r){2-4}
    \cmidrule(r){5-7}
    H Matrix (sparsity)  &\#unitaries& decomp time & coeff time & \#unitaries & decomp time & coeff time  \\
    \midrule
    32x32 (12.5\%)       & 63        & 0.0104      & 0.00024    & 63 (5)      &  0.006      & 0.0001      \\
    128x128 (3.51\%)     & 319       & 0.8882      & 0.0156    & 319 (7)      &  0.047      & 0.0016      \\
    512x512 (0.93\%)     & 1535      & 106.1       & 0.592     & 1,535 (9)    &  3.211      & 0.0548      \\
    2,048x2,048 (0.24\%) & -         & 63,375*     & -         & 7,167 (11)   &  111.4      & 0.9423      \\
    8,192x8,192 (0.06\%) & -         & -           & -         & 32,767 (13)  &  6,713      & 17.76      \\
    \bottomrule
  \end{tabular}
  \caption{\centering Comparison of decomposition and coefficient re-evaluation times or the matrices
                  in \Cref{tab-pc-sparsity}. Hadamard unitaries column includes \#clusters in brackets.
                  All times in seconds. * Estimate based on time to complete 10\% of products.}
  \label{tab-app-decomp-times}
\end{table}

Comparing the execution times, \Cref{tab-app-decomp-times} shows significant benefits for the
Hadamard approach for both the initial decomposition and the coefficient re-evaluation.
Even on a modest $512 \times 512$ matrix, the trace orthogonality approach is 30 times slower for the
initial decomposition and 10 times slower for the coefficient re-evaluation.
On the $2048 \times 2048$ matrix, the trace approach is estimated to require 17 hours based on the
time to complete 10\% of the trace orthogonality checks. This is nearly 600 times slower than the 
Hadamard approach. Since the trace based decomposition did not run to completion, the coefficient
evaluation time could not be evaluated.

The Hadamard approach creates clusters which include all Pauli strings that share the
cluster's sparsity pattern. Some clusters contain products that always have zero coefficients.
For example, for the $512 \times 512$ matrix, the 9 clusters contain a total of 2,304 Pauli strings
of which 769 always have a zero coefficient. The processing of these is included in the timings
in \Cref{tab-app-decomp-times} although only those unitaries with a non-zero coefficient are
included in the number of unitaries.
The zero coefficients arise because, whilst the PC matrix is non-symmetric, if does have a large number
of symmetric entries due to the finite volume discretisation that has been used.

Although the Hadamard approach has less severe scaling than the trace approach,
it took almost 2 hours to complete the LCU for the $8,192 \times 8,192$ matrix.
This corresponds to a $65 \times 65$ CFD mesh which is the level needed to to investigate turbulence
phenomena. 
This is a mesh of 4,225 nodes which is several orders of magnitude below the meshes
discussed at the start of \Cref{sec-cfd-alg}; and, far below the 166 million element mesh
at which \cite{scherer2017concrete} estimated quantum advantage would be achieved.
Although the LCU decomposition can be efficiently parallelised, it is likely to remain a 
significant bottleneck in hybrid CFD algorithms.

\begin{figure}[h]
  \begin{center}
    \includegraphics[width=0.90\textwidth]{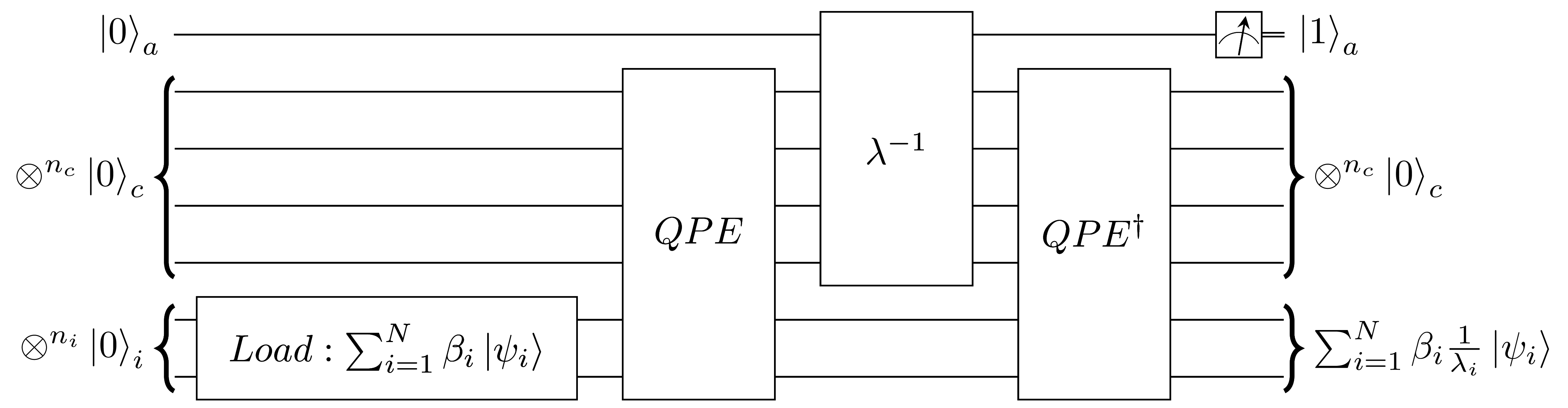}
  \end{center}
  \caption{\centering Outline of HHL algorithm \cite{harrow2009quantum}.}
  \label{fig-hhl-01}
\end{figure}

%
\subsection{Matrix solution}
The focus of this analysis is to demonstrate how the methodology shown in \Cref{fig-hybrid-cfd}
can be used to investigate the classical-quantum interface in a hybrid CFD solver.
For this, a completely error free quantum computer is emulated with a full state
loader as the input to HHL \cite{harrow2009quantum} and the full state vector as output from HHL.
Whilst this level of state preparation and measurement (SPAM) cannot be achieved in
practice it does allow some of the circuit parameters to be investigated.

The HHL algorithm \cite{harrow2009quantum} has been widely reported elsewhere and is not not repeated here.
The implementation details of the state vector simulator are:

\begin{itemize}
  \item{} {\bf State preparation.} In emulation mode, it is enough to directly write
          the input vector into the amplitudes of the state vector. However, this is not a 
          a framework for future study. Instead, the binary tree loader reported as
          algorithm 1 by \cite{araujo2021divide} is used. This creates a sequence
          of controlled one-qubit $N-1$ rotations for an input vector of rank $N$. \cite{mottonen2004transformation}.
  \item{} {\bf Matrix preparation.} The linear combination of unitaries are created using the 
          procedure described in \Cref{subsec-app-pauli-hadamard}. The exponentiation of the sum of 
          unitaries is approximated using the Trotter product formula \cite{cohen1982eigenvalue}.
          This creates a single unitary $U$ to be used in the phase estimation step.
  \item{} {\bf Phase Estimation.} The powers of $U$ applied to the input register are
          computed using a simple recursion: $U^{2n} \gets U^nU^n$. No circuit considerations are taken 
          into account when emulating the phase estimation step or the previous Trotterisation step.
  \item{} {\bf Eigenvalue Inversion.} The Polynomial State Preparation (PSP) of
           \cite{vazquez2018quantum, vazquez2020enhancing}
           is used for the simulated eigenvalue inversion circuit. The inversion function
           $\sin^{-1}\left( \frac{1}{x} \right)$
           is directly computed rather than estimated by a truncated Taylor series polynomial.
  \item{} {\bf Ancilla measurement.} For simulation, the ancilla qubit is directly projected onto
          the $\ket{1}$ state. Since the simulation has full access to the state vector, the expectation, $E$,
          of the $\ket{1}$ state being measured can be directly computed.
  \item{} {\bf Dimensional state vector.} The normalised solution, $\ket{\hat{x}}$, generated by HHL is 
          re-dimensionalised by:
\begin{eqnarray}
  \ket{x} = ||b||
  \frac{\sqrt{E}}{C}\ket{\hat{x}}
  \label{eqn-hhl-soln}
\end{eqnarray}
          Whilst $E$ is directly accessible in a simulation, it is a measured quantity on a physical
          device. Although it has not been tested in this work, marginal probabilities derived from the 
          state vector allow rapid emulation of a very large number of physical measurements, albeit 
          within the limits of classical pseudo random number generators. Accurate re-dimensionalisation 
          of the measured quantum state is expected to be an important factor in hybrid CFD algorithms.
\end{itemize}

As shown in \Cref{fig-hybrid-cfd}, the full (simulated) state vector is returned to the hybrid interface.

%
\section{Results}
\label{sec-results}
All results were run on a desktop PC with an 
Intel\textsuperscript{\tiny\textregistered}
Core\textsuperscript{\tiny\textcopyright}
i9 12900K 3.2GHz
Alder Lake processor and 64GB of DDR4 RAM. 
All calculations were run in serial mode.
Whilst serial execution may affect the precise performance results, 
the relative comparisons
between the unitary decomposition and the CFD solver run times are broadly valid.
However, the timings should only be taken as indicative as only 
preliminary code optimisation has
been performed for both the CFD solver and the hybrid interface. All performance comparisons exclude the time taken to simulate HHL.

In all tests, the matrices created by summing the unitaries were compared with the original CFD
matrices and, in all cases, the differences were $\mathcal O(10^{-15})$ or less and, therefore,
within the rounding error of the double precision arithmetic used in the algorithm.
This section first analyses the unitary decompositions that are produced.
The decompositions are then used as the basis for simulated HHL solutions. 
Firstly, the individual pressure correction matrices sampled from the classical CFD run are
considered.
Secondly, a full hybrid run of the CFD solver is performed with each pressure correction step being
solved using the unitary coefficient update procedure and the HHL algorithm.

\begin{table}[h]
  \centering
  \begin{tabular}{c c c c c c c}
    \toprule
    CFD mesh &  \#unitaries & \% non-zeros & decomp time & coeff time & CFD time & Hybrid O/H\\
    \midrule
     5x5     & 63          & 50.0\%       &  0.006      & 0.0001     &  0.0085 & 4.8  \\
     9x9     & 319         & 55.6\%       &  0.047      & 0.0016     &  0.056  & 8.1  \\
    17x17    & 1,535       & 62.7\%       &  3.211      & 0.0548     &  1.127  & 15.0 \\
    33x33    & 7,167       & 71.8\%       &  111.4      & 0.9423     &  12.13  & 28.6 \\
    65x65    & 32,767      & 81.0\%       &  6,713      & 17.76      &  210.1  & 53.1 \\
    \bottomrule
  \end{tabular}
  \caption{\centering Comparison of decomposition and coefficient re-evaluation times for the matrices
                  in \Cref{tab-pc-sparsity} including comparison with the CFD solver time.
                  All times in seconds.}
  \label{tab-pc-times}
\end{table}

%
\subsection{Unitary decomposition and coefficient update}

\Cref{tab-pc-times} shows the unitary decomposition times for the range of CFD meshes
using the pressure correction equation after 10 outer iterations.
The corresponding eigenvalues and condition numbers were given in \Cref{tab-pc-sparsity}.
The number of unitaries as a percentage of the number of non-zeros in the matrix shows
that the number of unitaries is rising slightly faster than $\mathcal O(N)$.
The unitary decomposition times from \Cref{tab-app-decomp-times} are compared with
the time taken for a complete classical CFD solution.
On all meshes, the one-off initial unitary decomposition time is roughly equal to or much greater
than the time for the classical CFD solution.
The column labelled 'Hybrid O/H' is the accumulated time of the unitary decomposition
(one-off and repeated coefficient evaluations) divided by the time to run the CFD solver.
This roughly doubles each time the mesh size increases to the point that the time 
spent in the hybrid interface on the 65x65 mesh is 53 times that of a classical CFD solution.
All the timings are purely for those hybrid tasks that must be performed on a 
classical computer and exclude the time to emulate HHL.

Considering the finite element mesh of \cite{scherer2017concrete} 
with 332,020,680 edges, each edge corresponds to 2 entries in the assembled matrix.
The scaling of \Cref{tab-pc-times} suggests that the corresponding LCU would 
consist of the order of one billion unitaries if this were a CFD mesh.

\begin{figure}[h]
  \centering
  \begin{subfigure}[b]{0.45\textwidth}
      \centering
      \includegraphics[width=\textwidth]{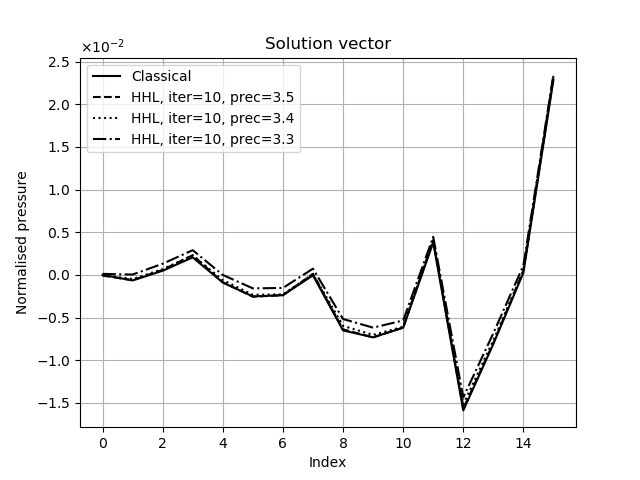}
      \caption{\small Solution for iteration 10 matrix.}
      \label{fig-4x4-hhl-iter10}
  \end{subfigure}
  \hfill
  \begin{subfigure}[b]{0.45\textwidth}
      \centering
      \includegraphics[width=\textwidth]{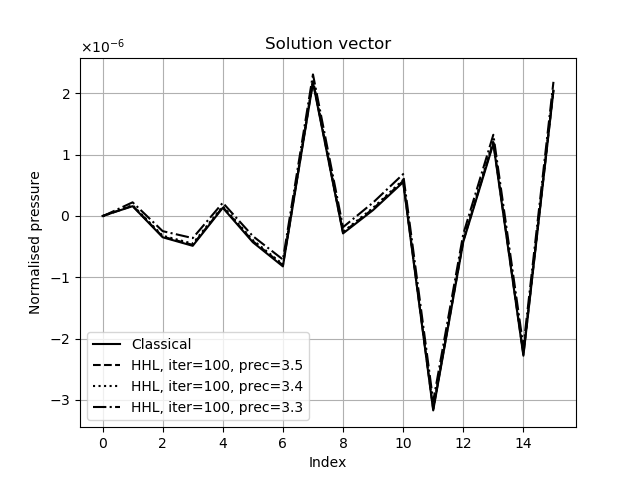}
      \caption{\small Solution for iteration 100 matrix.}
      \label{fig-4x4-hhl-iter100}
  \end{subfigure}
  \caption{\centering HHL solutions for the $5 \times 5$ mesh saved after 10 and 100 outer iterations.
                  The precision $m.n$ indicates $m$ integer qubits and $n$ fraction qubits. All
                  cases have a sign qubit, meaning the clock register contains $m+n+1$ qubits.}
  \label{fig-hhl-prec}
\end{figure}

%
\subsection{HHL solutions for sampled pressure correction matrices}

\Cref{fig-hhl-prec} compares the HHL solutions for the iteration 10 and 100
pressure correction matrices and input
vectors shown in \Cref{fig-cavity-bx}.
For the HHL solutions, three precision settings are considered for the
QPE clock register. Since the symmetrised matrix, $H$, has negative eigenvalues, all
calculations require a sign qubit. 
The precisions considered give the following ranges and numbers of qubits:

\begin{itemize}
  \item{} $3.3$ - range $[\pm 0.125, \pm 8.875]$ using 7 clock qubits and 13 qubits in total.
  \item{} $3.4$ - range $[\pm 0.0625, \pm 8.9375]$ using 8 clock qubits and 14 qubits in total.
  \item{} $3.5$ - range $[\pm 0.03125,\pm 8.96875]$ using 9 clock qubits and 15 qubits in total.
\end{itemize}

The total number of qubits includes 5 for the input register and 1 for the ancilla qubit.
Comparing with the eigenvalues from \Cref{tab-pc-sparsity} it can be seen that all precisions
cover the largest eigenvalue of $4.54$, but only the last one covers the lowest eigenvalue
of $5.2 \times 10^{-2}$.

\Cref{fig-4x4-hhl-iter10} shows that best precision ($3.5$) matches almost exactly the
classical solution. The $3.4$ precision HHL solution is also very close to the the classical solution.
At the $3.3$ precision, the HHL solution is starting to degrade.
\Cref{fig-4x4-hhl-iter100} shows that the simulated HHL performs well even for input states with a
norm close to the single precision limit for real numbers.
\Cref{tab-5x5-prec-fid} shows the influence of the precision on the fidelity between the HHL and exact
solutions.
Of interest is the fact that the fidelity does not degrade as the magnitude of the input vector
reduces between 10 and 100 iterations. However, the expectation of the ancilla qubit being in
the $\ket{1}$ state does reduce from $0.6\%$ to $0.05\%$.

\begin{table}[h]
  \centering
  \begin{tabular}{c c c }
    \toprule
    Precision & 10 iterations  & 100 iterations  \\
    \midrule
     3.3      &   0.99342      & 0.99593         \\ 
     3.4      &   0.99934      & 0.99963         \\
     3.5      &   0.99977      & 0.99991         \\
    \bottomrule
  \end{tabular}
  \caption{\centering Fidelity between HHL solution and exact solution on sample matrices for $5 \times 5$ mesh.}
  \label{tab-5x5-prec-fid}
\end{table}

Based on these results, the $3.4$ precision option is used for the subsequent simulations on the
$5 \times 5$ mesh.

%
\subsection{Hybrid HHL solutions for cavity test case}
\begin{figure}[h]
  \centering
  \begin{subfigure}[b]{0.45\textwidth}
      \centering
      \includegraphics[width=\textwidth]{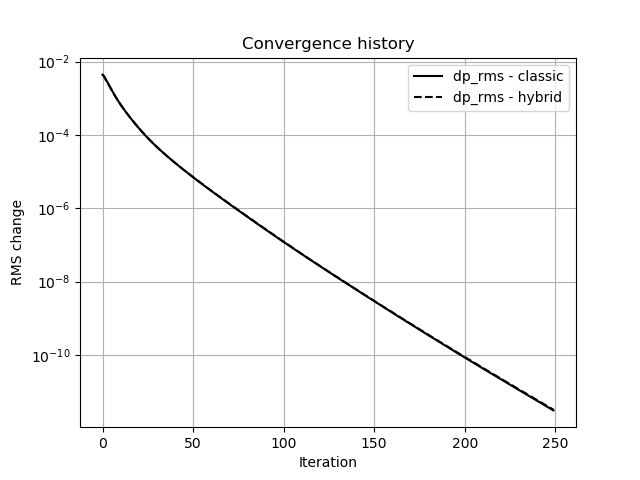}
      \caption{\small Convergence of pressure updates.}
      \label{fig-4x4-hhl-dp-hist}
  \end{subfigure}
  \hfill
  \begin{subfigure}[b]{0.45\textwidth}
      \centering
      \includegraphics[width=\textwidth]{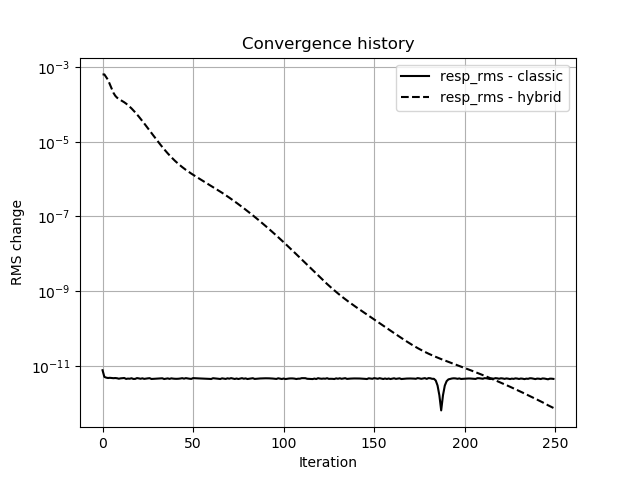}
      \caption{\small Convergence of continuity equation residuals.}
      \label{fig-4x4-hhl-resp-hist}
  \end{subfigure}
  \caption{\centering Comparison of classical and hybrid convergence histories for $5 \times 5$ mesh.}
  \label{fig-hhl-hist}
\end{figure}

\Cref{fig-hhl-hist} compares the convergence histories of the classical
and hybrid solutions with both terminated after 250 outer iterations for the $5 \times 5$ CFD mesh.
The graphs show the two measures of convergence of the pressure correction equation:
the root mean square of the pressure correction and the root mean square mass conservation
residual. The convergence histories for the u and v velocities have the same character as
the pressure correction convergence.

For the convergence of the pressure correction update, \Cref{fig-4x4-hhl-dp-hist},
the classical and hybrid schemes are essentially identical.
More interesting is the convergence of the mass conservation residual in \Cref{fig-4x4-hhl-resp-hist}.
A feature of the SIMPLE scheme is that the pressure correction equation is linear and,
hence, the corrected velocities should satisfy the continuity equation to within machine 
precision. The classical solution achieves this. 
The hybrid solution does not, which indicates that some of the finer details of the
pressure correction solution are lost due to the precision of the clock register.
Nevertheless, the hybrid scheme is able to converge the continuity residual to close
to machine precision by the $250^{th}$ iteration.
The reason that an HHL precision of $3.4$ can achieve a residual precision 
of $\mathcal O(10^{-12})$ is the normalisation of the input vector.
The differences in the continuity residual do not affect the momentum equations for which
the match between the classical and hybrid solutions is the same as for the pressure
corrections.
These results were somewhat unexpected as it was anticipated that the reduced precision of the
HHL clock register would have a larger impact.

%
\subsubsection{Partial unitary decompositions}
\label{subsubsec-partial-LCU}
To illustrate the use of a simulated hybrid framework, the influence of ignoring
unitaries in the LCU that have small coefficients is considered.
\Cref{tab-coeff-limit} shows the impact of increasing this threshold on the number of unitaries in
the LCU using the sampled pressure correction matrix after 10 outer iterations.
The limits were chosen heuristically based on roughly halving the number of unitaries for
$|\alpha|>4 \times 10^{-02}$ and halving again for $|\alpha|>5 \times 10^{-02}$.

\begin{table}[h]
  \centering
  \begin{tabular}{c c}
    \toprule
    LCU coefficient limit             & number of unitaries \\
    \midrule
     $1 \times 10^{-06}$   &  63   \\
     $1 \times 10^{-02}$   &  53   \\
     $2 \times 10^{-02}$   &  48   \\
     $3 \times 10^{-02}$   &  46   \\
     $4 \times 10^{-02}$   &  33   \\
     $5 \times 10^{-02}$   &  14   \\
     $1 \times 10^{-01}$   &  10   \\
    \bottomrule
  \end{tabular}
  \caption{\centering Influence of ignoring small coefficients on number of unitaries in LCU, 
                  sample pressure correction matrix after 10 iterations. Coefficients with
                  $|\alpha|$ less than the limit are ignored.}
  \label{tab-coeff-limit}
\end{table}

\begin{figure}[h]
  \centering
  \begin{subfigure}[b]{0.45\textwidth}
      \centering
      \includegraphics[width=\textwidth]{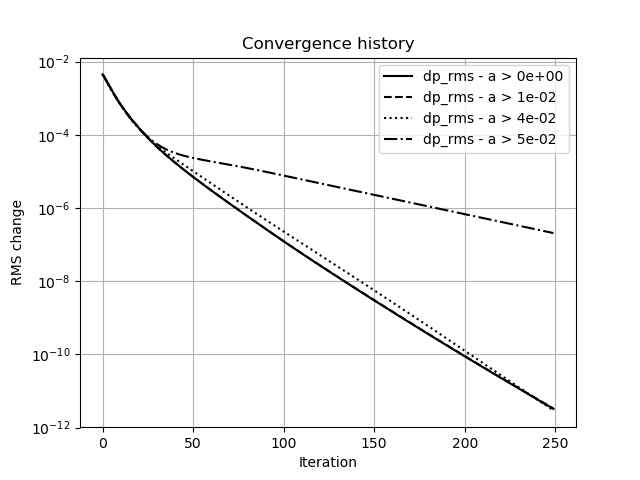}
      \caption{\small Convergence of pressure updates.}
      \label{fig-4x4-hhl-dp-comp}
  \end{subfigure}
  \hfill
  \begin{subfigure}[b]{0.45\textwidth}
      \centering
      \includegraphics[width=\textwidth]{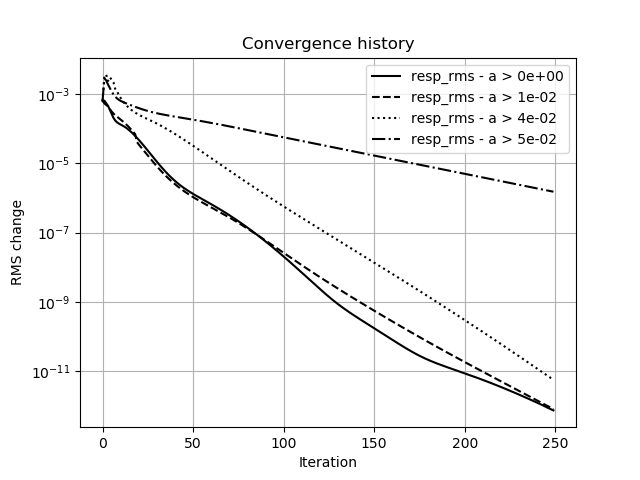}
      \caption{\small Convergence of continuity equation residuals.}
      \label{fig-4x4-hhl-resp-comp}
  \end{subfigure}
  \caption{\centering Influence of ignoring small LCD coefficients on convergence of the hybrid CFD solver.}
  \label{fig-hhl-comp}
\end{figure}

\Cref{fig-hhl-comp} shows the influence of ignoring small LCD coefficients on the convergence 
of the hybrid CFD solver.
If the limit is set to $1 \times 10^{-02}$ approximately $15\%$ of the coefficients are ignored
and, other than small impact on the continuity residual, the impact is negligible.
Increasing the limit to $4 \times 10^{-02}$ omits just under $50\%$ of the coefficients and
whilst the impact is more noticeable, the rate of convergence is only marginally impacted.
Increasing the limit further to  $5 \times 10^{-02}$ now omits over $75\%$ of the coefficients 
and markedly degrades the rate of convergence.
However, the results for the $5 \times 10^{-02}$ limit do suggest that the calculation
would eventually converge.
As is often observed in CFD, the boundary between good and poor behaviour is quite narrow
and thresholds are usually case dependent.
An approach based on omitting the $50\%$ of unitaries that have the lowest coefficients
may enhance the performance of the quantum circuit but does not help the hybrid interface
as this is an iteration by iteration assessment that requires the full unitary
decomposition.

%
\subsubsection{Eigenvalue inversion}
\label{subsubsec-eval-inv}

The reason that \Cref{tab-5x5-prec-fid} shows such high fidelities for the HHL solutions is 
that the eigenvalue inversion circuit has used a full bit-wise accurate circuit.
In the terminology of \cite{vazquez2018quantum}, the eigenvalue inversion circuit as the same 
degree has the number of clock qubits.

\begin{figure}[h]
  \begin{center}
    \includegraphics[width=0.90\textwidth]{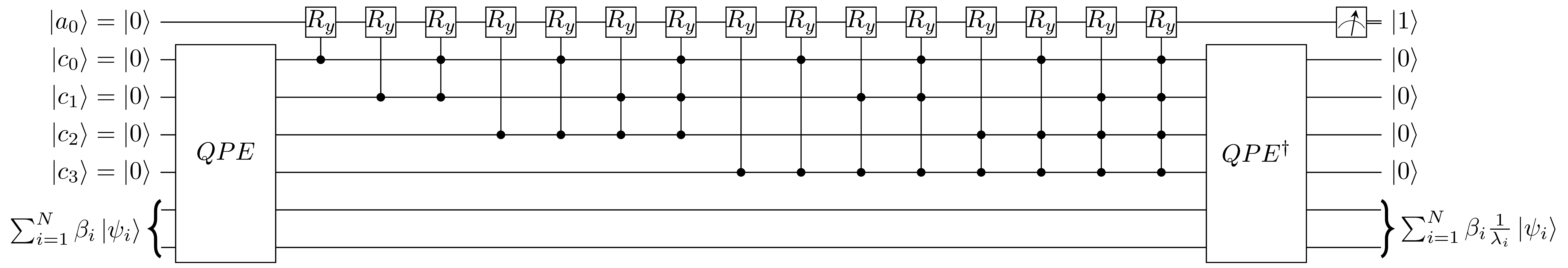}
  \end{center}
  \caption{\centering Sample HHL circuit with full degree eigenvalue inversion circuit.}
  \label{fig-evali-circ}
\end{figure}

\Cref{fig-evali-circ} illustrates such a full degree inversion for a circuit with 4 clock
qubits. As can be seen, all possible bit patterns are used to control separate rotations of
the ancilla qubit. 
This leads to an inversion circuit of depth $2^{n_c}-1$.
For larger clock registers, the simulation time is dominated by the eigenvalue inversion.
Since this is an $\mathcal O(N)$ circuit it is also likely to scale poorly on physical 
devices. It is, therefore, instructive to consider the accuracy of the eigenvalue 
inversion circuit on the CFD solution.

One of the rationales of polynomial state preparation \cite{vazquez2018quantum} is that 
smooth functions can be approximated by low degree polynomials and hence shallower circuits.
However, the state of the clock register after QPE is likely to be highly non smooth with
peaks corresponding to the eigenvalues and troughs elsewhere.
With a precision of $\pm3.4$ for the $5 \times 5$ mesh, the full eigenvalue inversion circuit
is of order 8 and contains 255 ancilla rotations giving a fidelity of 0.99963 
from \Cref{tab-5x5-prec-fid}.
Reducing the order to 7, saves just one rotation and reduces the fidelity to 0.98841.
Reducing the order to 6, saves 9 rotations and reduces the fidelity to 0.88418.
As expected, reducing the order of the circuit strongly affects the solution and 
saves hardly any rotations.

An alternative is omit those bit-patterns in the PSP that have a low probability after 
phase estimation.
This is easy to implement in a simulation but would be hard on a physical device. 
After QPE the state space amplitudes contain the probabilities that each bit-pattern in
the clock register corresponds to an eigenvalue, although these are embedded with the 
state-space amplitudes for the entire circuit.
In emulation, marginal probabilities for the clock register can be easily
extracted from the state space amplitudes. These can be used to omit the
ancilla rotations for states that have a small marginal probability.

\Cref{tab-5x5-marg-prob} shows the effect of the probability limit on the number of
ancilla rotations and the fidelity for the $5 \times 5$ sample matrix after 10 and 100 outer iterations.
At 10 iterations, a limit of $1 \times 10^{-04}$ has a marginal effect on the fidelity and 
over half the ancilla rotations are saved.
At 100 iterations, a larger number of ancilla are saved and even at a limit
of $1 \times 10^{-03}$ a fidelity greater than 0.99 is achieved.

\begin{table}[h]
  \centering
  \begin{tabular}{c c c c c}
    \toprule
    \multicolumn{1}{c}{Probability} & \multicolumn{2}{c}{10 iterations} & \multicolumn{2}{c}{100 iterations} \\
    \cmidrule(r){2-3}
    \cmidrule(r){4-5}
    Limit & No. of ancilla rotations & Fidelity & No. of ancilla rotations & Fidelity   \\
    \midrule
     $1 \times 10^{-06}$   &   255     & 0.99593  &   255     & 0.99963  \\
     $1 \times 10^{-05}$   &   192     & 0.99336  &   156     & 0.99963  \\
     $1 \times 10^{-04}$   &   120     & 0.99278  &   80      & 0.99489  \\
     $1 \times 10^{-03}$   &   68      & 0.98574  &   36      & 0.99426  \\
    \bottomrule
  \end{tabular}
  \caption{\centering Influence of omitting ancilla rotations below the marginal probability limits on
                  the sample matrix after 10 and 100 iterations for $5 \times 5$ mesh.}
  \label{tab-5x5-marg-prob}
\end{table}

\begin{figure}[h]
  \centering
  \begin{subfigure}[b]{0.45\textwidth}
      \centering
      \includegraphics[width=\textwidth]{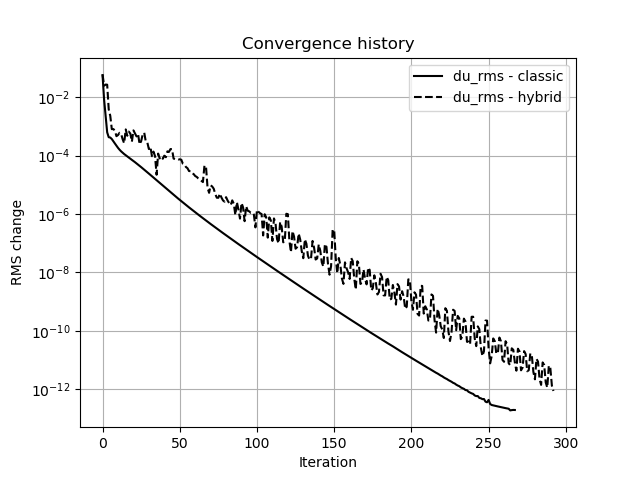}
      \caption{\small Convergence of $u$ velocity updates.}
      \label{fig-4x4-u-mpe-03}
  \end{subfigure}
  \hfill
  \begin{subfigure}[b]{0.45\textwidth}
      \centering
      \includegraphics[width=\textwidth]{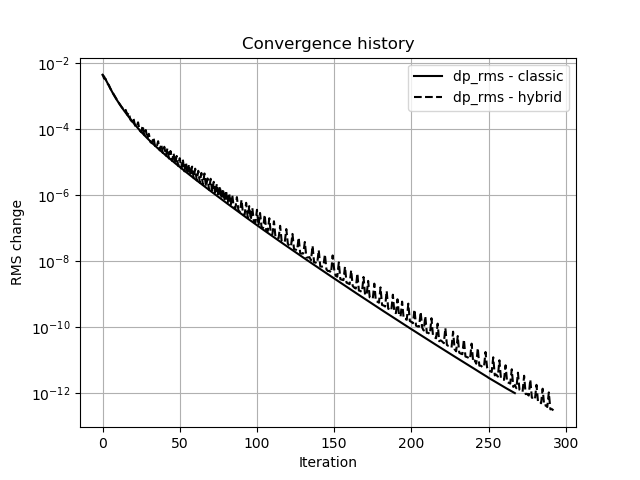}
      \caption{\small Convergence of pressure updates.}
      \label{fig-4x4-p-mpe-03}
  \end{subfigure}
  \caption{\centering Convergence histories on $5 \times 5$ mesh for eigenvalue inversion where ancilla rotations 
                  with a marginal probability  $<1 \times 10^{-03}$ are ignored.}
  \label{fig-4x4-mpe-03}
\end{figure}

\Cref{fig-4x4-mpe-03} shows the $u$-velocity and pressure convergence histories where 
ancilla rotations with a marginal probability  $<1 \times 10^{-03}$ are ignored.
Whilst the fidelities in \Cref{tab-5x5-marg-prob} suggest this limit should have a marginal
effect, there is a noticeable impact on convergence of the $u$-velocity.
The convergence also demonstrates a more oscillatory nature which is characteristic of
integer decisions (the number of rotations to ignore) which are based on real-valued
thresholds (the marginal probability).
However, the hybrid solution does still converge after 293 outer iterations compared to 268
for the classical solver. 
As in classical computing a scheme that requires more rapid iterations may 
out-perform a scheme that user fewer more complicated iterations. 
Unfortunately, this assessment can only be performed with a meaningful outcome 
on a physical device.

%
\subsection{9x9 mesh}
\label{subsec-9x9}
Results for the $9\times 9$ mesh are briefly considered in this section. 
The precision required to span the eigenvalue range shown in
\Cref{tab-pc-sparsity} is$\pm1.9$.
The smallest non-zero number that can be represented by this precision is
$1.95\times 10^{-03}$ which compares to the 
smallest eigenvalue from \Cref{tab-pc-sparsity} of $2.7\times 10^{-03}$.
\Cref{tab-9x9-marg-prob} combines the effects of precision and marginal probability for
the pressure correction matrix after 10 iterations for the $9 \times 9$ mesh.
As expected, it is only with a precision of $\pm1.9$ that a fidelity $>0.99$ is reached.
However, reducing the number of fraction qubits has a more marked effect than
for the $5 \times 5$ case shown in \Cref{tab-5x5-prec-fid}. 
In the $5 \times 5$ case, a 2 qubit reduction still produces a fidelity $>0.99$ whereas
for the $9 \times 9$ case the fidelity drops to $<0.95$. This is first indication
that results for small meshes may not extrapolate to larger ones.

\begin{table}[h]
  \centering
  \begin{tabular}{c c c c c c c}
    \toprule
    \multicolumn{1}{c}{Precision} & \multicolumn{3}{c}{Full circuit} & \multicolumn{3}{c}{Partial circuit} \\
    \cmidrule(r){2-4}
    \cmidrule(r){5-7}
               & No. of ancilla rotations & Fidelity & No. of ancilla rotations & Fidelity \\
    \midrule
     1.6    &   511     & 0.93846  &  511     & 0.93846   \\
     1.7    &   1,023   & 0.94503  &  594     & 0.94503   \\
     1.8    &   2,048   & 0.98447  &  616     & 0.98446   \\
     1.9    &   4,095   & 0.99967  &  982     & 0.99965   \\
    \bottomrule
  \end{tabular}
  \caption{\centering Influence of precision and omitting ancilla rotations below a marginal probability 
                  of $1\times 10^{-06}$ on the sample matrix after 10 iterations for the $9 \times 9$ mesh.}
  \label{tab-9x9-marg-prob}
\end{table}

\begin{figure}[h]
  \centering
  \begin{subfigure}[b]{0.45\textwidth}
      \centering
      \includegraphics[width=\textwidth]{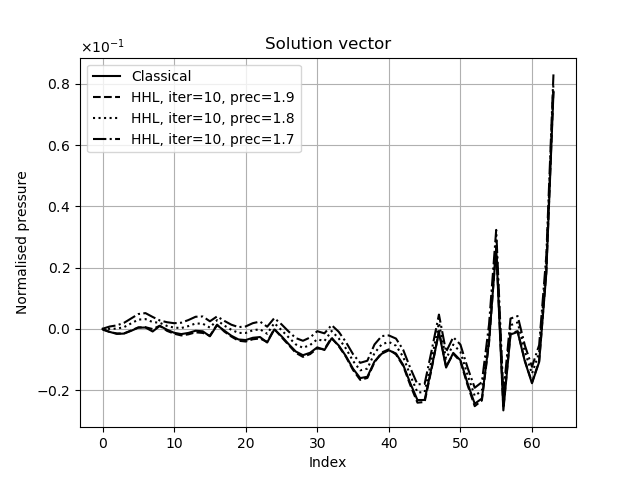}
      \caption{\small Solution for iteration 10 matrix.}
      \label{fig-8x8-iter10}
  \end{subfigure}
  \hfill
  \begin{subfigure}[b]{0.45\textwidth}
      \centering
      \includegraphics[width=\textwidth]{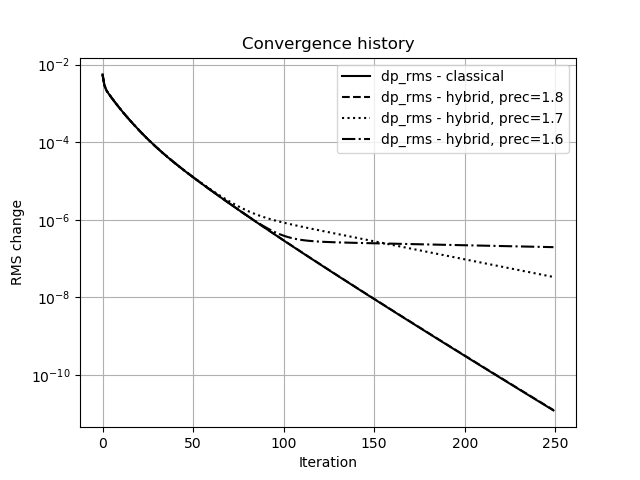}
      \caption{\small Convergence of pressure updates.}
      \label{fig-8x8-conv}
  \end{subfigure}
  \caption{\centering Comparison of classical and HHL solutions for the $9 \times 9$ mesh matrix after
                  10 iterations; and, the full CFD solution using HHL with precision
                  $\pm2.6$ for 100 iterations.}
  \label{fig-8x8-comp}
\end{figure}

\Cref{fig-8x8-iter10} compares the HHL solutions at the listed precisions with the classical
solution for a sampled matrix after 10 iterations from the $9 \times 9$ mesh. 
The classical and $\pm1.9$ precision solutions are identical as expected from the 
fidelity in \Cref{tab-5x5-marg-prob}. The lower precisions show the same character as
the classical solution but are offset and in some of the smaller values have the opposite sign.
The $\pm1.6$ solution is omitted as it is very close to the $\pm1.7$ solution.

\Cref{fig-8x8-conv} compares the full CFD convergence history for the classical and hybrid solutions. The $\pm1.9$ precision hybrid case was not run as the $\pm1.8$ precision
solution is virtually identical to the classical solution. 
The impact of of the drop in fidelity in \Cref{tab-9x9-marg-prob} is
seen in the convergence histories for the $\pm1.6$ and $\pm1.7$ solutions.
The latter does appear as though it would ultimately converge, but
the $\pm1.6$ convergence has stagnated.
In general, the larger eigenvalues, which are all retained, correspond to the
larger scale flow structures. The smaller eigenvalues correspond to structures
that vary between mesh cells. Omitting these appears to prevent the cell-wise residuals
being reduced to machine precision.

In CFD literature, it is common to express errors in terms of their
root mean square (RMS):

\begin{equation}
  e_{rms} = \sqrt{\frac{1}{N}\sum_{k=1}^{N} \lvert\delta x_k\rvert^2}
  \label{eqn-evals-rms}
\end{equation}

\Cref{tab-9x9-fid-err} lists fidelities and error norms for the
solution of the matrix after 10 iterations. This suggests that
the RMS error cannot be much above $1\%$ before the hybrid 
algorithm struggles to converge. This is likely to have
implications for the fidelity with which the {\it pseudo} state vector is
reconstructed from measurements on a physical quantum computer.

\begin{table}[h]
  \centering
  \begin{tabular}{c c c c }
    \toprule
    Precision & Fidelity & $L_2$ error norm & RMS error  \\
    \midrule
     1.6      &   0.9385      & 0.3507    & 0.0310    \\ 
     1.7      &   0.9450      & 0.3319    & 0.0293    \\
     1.8      &   0.9848      & 0.1721    & 0.0154    \\
     1.9      &   0.9997      & 0.0245    & 0.0022    \\
    \bottomrule
  \end{tabular}
  \caption{\centering Fidelity and error norms between HHL solution and exact solution on
  for sample matrix after 10 iterations for the $9 \times 9$ mesh.}
  \label{tab-9x9-fid-err}
\end{table}

The foregoing analysis of the results has demonstrated that with fidelities 
greater than 0.98, a hybrid HHL solution can produce results of equal quality to the
classical solution. There are some options for reducing the number of unitaries in the LCU
and for reducing the number of rotations of the ancilla qubit in the eigenvalue circuit.
It is recognised that the later would be a challenge to implement on a physical device.
In most cases, the reductions have maintained fidelities above or close to 0.99 for the
match between the HHL solution and the classical one.
Below this, the $9 \times 9$ mesh solutions have shown that the quality 
of the inner QLES solution affects the convergence of the outer CFD solution. 
This can be just a reduced rate of convergence or a stagnation of convergence.

%
\section{Conclusions}
\label{sec-conclusion}
A concern for future quantum advantage for CFD applications is the computational effort 
required on the classical side of a hybrid interface. This could lie in classical costs
associated with data preparation or circuit design or some other factor yet to emerge.
It is certainly not axiomatic that the computational cost of the hybrid interface will be
lower than the cost of performing the equivalent classical CFD calculation; or,
that making it so does not incur significant classical supercomputing costs.

An emulation of a hybrid CFD solver using HHL on the smallest $5 \times 5$ mesh has 
produced benchmark results showing that HHL does not need full 64-bit precision to 
match the iterative performance of the CFD solver and can achieve convergence levels of
$\mathcal O(10^{-12})$. However, preliminary timing results suggest that the classical
side of the hybrid interface causes the hybrid solver to take 10 times longer than
the classical solver.

An investigation of the number of unitaries in the LCU for CFD matrices up to
$65 \times 65$ shows that the number of unitaries scales slightly faster than 
linearly in the number of non-zeros in the CFD matrix, with the largest
matrix requiring 32,767 Pauli strings.
A new Hadamard based orthogonality approach for finding the Pauli strings in a LCU
outperforms the trace orthogonality approach on the test matrices.
However, it only marginally delays the inevitable impact of the $>\mathcal O(N^2)$ scaling.
For the $65 \times 65$, finding the Pauli strings in the LCU decomposition alone
takes 30 times longer than the full CFD solver. 
The Pauli string algorithm is more easily parallelised than a CFD solver but the cross-over
point for industrial scale CFD applications is likely to require significant super-computing
costs.
Note that classical parallelisation does not reduce compute resources, it speeds up 
execution by amassing the resources of multiple devices.

An analysis based on omitting unitaries with small coefficients has shown that up to half the
unitaries can be omitted in the QLES without materially affecting the convergence of the hybrid solver.
However, there is a small margin before further omissions degrade the performance.
Dynamic algorithms that omit $50\%$ of unitaries with the lowest coefficients would
aid the quantum circuit but not the classical costs of the hybrid interface.

A study of the eigenvalue inversion circuit has shown that a full order circuit
is needed to match the classical solution. Using the fact that the emulation has
full access to the state vector after the QPE step has shown, the unsurprising result,
that ancilla rotations for eigenvalues with low probabilities can be ignored.
For the $9 \times 9$ mesh this corresponded to a saving of over 75\%. However, this is 
not considered a practical alternative for real devices. 
For the largest $65 \times 65$ mesh, the number
of clock qubits for QPE is estimated to be 20 giving a full order eigenvalue inversion
circuit with 1,048,576 ancilla rotations. 
It does not seem unreasonable to expect
that industrial CFD applications will require clock registers with at least 32 qubits
and possibly 64 qubits with the latter requiring a full eigenvalue inversion circuit
with $\mathcal O(10^{19})$ ancilla rotations.

This paper began by comparing the current state of quantum CFD with that of classical CFD in 
the 1970s. The intervening years have taught us that performance bottlenecks can be overcome.
There are other quantum algorithms for solving linearised equations, most notably 
Quantum Singular Value Transformation (QSVT). These are equally likely to require 
classical pre-processing whose compute costs scale with the problem size at faster rate than
the quantum solver. In the case of QSVT, these are likely to be the costs of 
calculating  $\kappa$ and the phase factors for the rotations of the signal qubit.
Non-linear quantum CFD algorithms are likely to emerge which will reduce or eliminate the
iteration to iteration classical pre-processing, but the effort required by the initial one-off 
classical pre-processing is unclear and, as shown, this can be dominant.
Whatever the algorithm, quantum advantage for industrial scale CFD is very likely to be reliant
on classical supercomputing. The question is how much?

Finally, this work has sought to provide CFD test cases and benchmark solutions 
that can be feasibly run on the first generation of fault tolerant devices.

%
\section{Data availability}
The sample matrices and vectors for the meshes listed in \Cref{tab-pc-sparsity}
after 10 and 100 iterations are available from the author upon request.

%
\section{Acknowledgements}
The permission of Rolls-Royce to publish this paper is gratefully acknowledged.
The HHL results were completed as part of funding received under the UK's
Commercialising Quantum Technologies Programme (Grant reference 10004857).
This work has benefited from several technical discussions and the author would like 
to thank Neil Gillespie, Joan Camps and Christoph Sunderhauf of Riverlane; and,
Jarrett Smalley of Rolls-Royce.

%
\appendix
%
\section{The SIMPLE CFD algorithm}
\label{sec-app-cfd-alg}

The objective of this appendix is to provide sufficient detail of Computational Fluid Dynamics (CFD)
to describe the context in which quantum algorithms may be applied.
The descriptions are intended to give a high level overview to those with little prior knowledge of CFD.
Whilst there are a range of CFD algorithms and modelling choices, attention is restricted to
the SIMPLE (Semi-Implicit Method for Pressure Linked Equations) algorithm 
\cite{patankar1972calculation}, \cite{ghia1977study}.
Derivatives of this algorithm are still widely used in modern industrial CFD codes, e.g. \cite{ammour2018subgrid}.

To simplify the derivation, consider the steady 2-dimensional form of the
Navier-Stokes equations \Crefrange{eqn-navier-rho}{eqn-navier-u} which can be written in component
form as:

\begin{equation}
  \begin{array}{rcl}
    \frac{\partial \rho u}{\partial x} + \frac{\partial \rho v}{\partial y} &=& 0 \\
    \frac{\partial \rho uu}{\partial x} + \frac{\partial \rho vu}{\partial y} &=& - \frac{\partial p}{\partial x} +
          \frac{\partial}{\partial x} \left( \mu \frac{\partial u}{\partial x}\right) +
          \frac{\partial}{\partial y} \left( \mu \frac{\partial u}{\partial y}\right) \\
    \frac{\partial \rho uv}{\partial x} + \frac{\partial \rho vv}{\partial y} &=& - \frac{\partial p}{\partial y} +
          \frac{\partial}{\partial x} \left( \mu \frac{\partial v}{\partial x}\right) +
          \frac{\partial}{\partial y} \left( \mu \frac{\partial v}{\partial y}\right) \\
  \end{array}
  \label{eqn-navier-2d}
\end{equation}

%
\subsection{Discretisation}
Analytical solutions to the Navier-Stokes equations exists in only the rarest cases and
practical applications of CFD require a numerical solution procedure.
The equations are solved over a region of interest with specified conditions applied at
the boundaries of the region.

\begin{figure}[h]
  \begin{center}
    \includegraphics[width=0.50\textwidth]{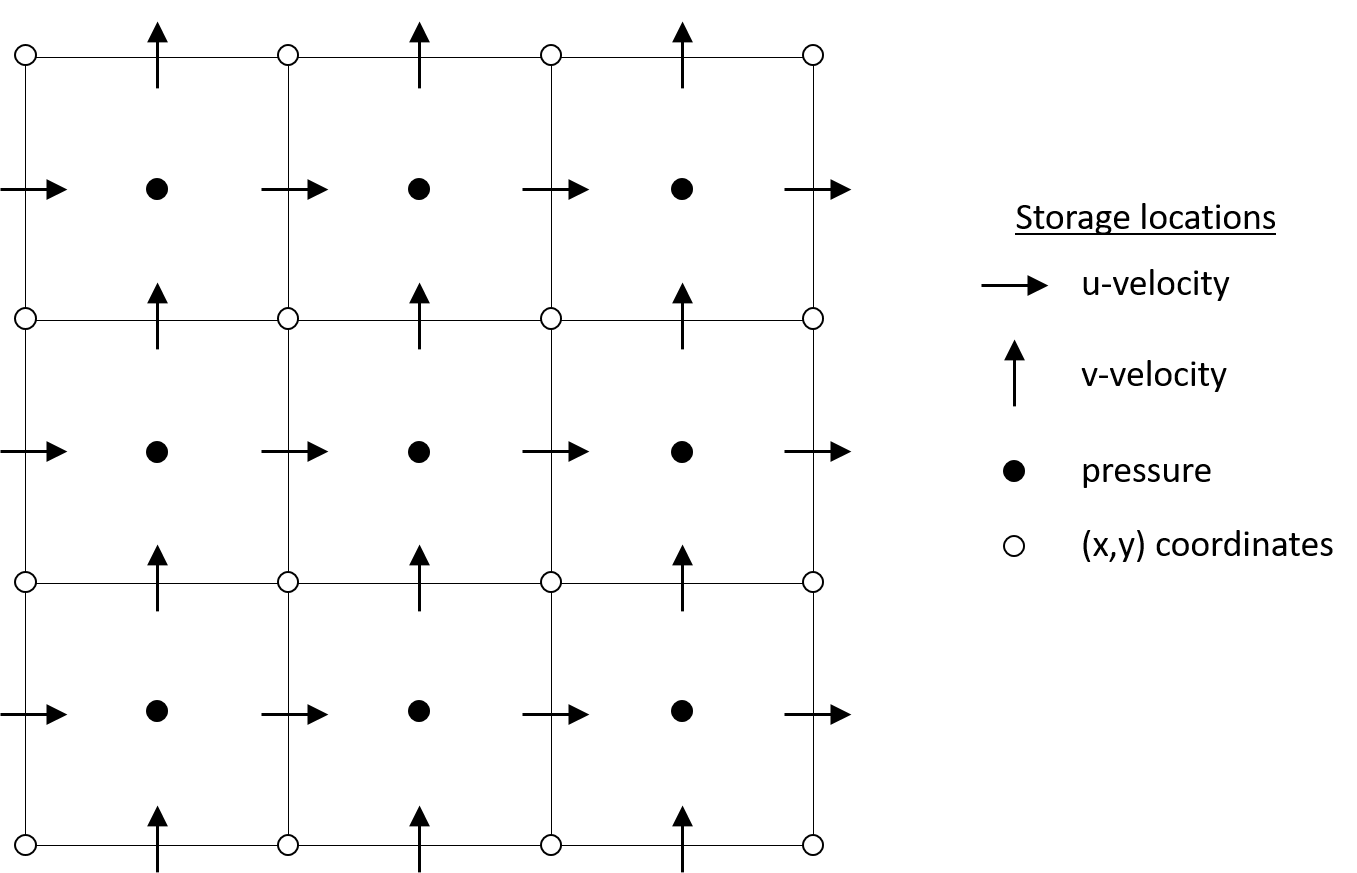}
  \end{center}
  \caption{\small Staggered mesh arrangement for SIMPLE pressure correction solver.}
  \label{fig-staggered-grid-01}
\end{figure}

In order to solve the equations numerically, the continuous flow variables are represented
by a discrete set of values, e.g.
\begin{equation}
  u_{i,j} \simeq  u(x_i, y_j), \quad i=1,n, \quad j=1,m
  \label{eqn-uij}
\end{equation}

where $(x_i, y_j) \in \mathcal M_u \subset \mathbb{R}^2$.
$\mathcal M_u$ is the {\it mesh} that contains the coordinates of all the points, usually called nodes,
at which discrete values of $u$ will be computed. There are two important properties of this structure:
\begin{itemize}
  \item{} The mesh coordinates can be referenced independently by indices $i$ and $j$ in the
          $x$ and $y$ directions.
          This means that the mesh has a lattice structure. In general, the lattice can be
          stretched and distorted, as is needed, for example, to model the flow over a curved
          body such as an airfoil. For this study, a simple Cartesian lattice is sufficient.
  \item{} The mesh coordinates for each flow variable do not need to be the same.
          Here, the {\it staggered} mesh arrangement of \cite{harlow1965numerical}
          and \cite{patankar1972calculation}, is used in which there are separate meshes
          $\mathcal M_u$, $\mathcal M_v$ and $\mathcal M_p$ for the discrete values of
          $u$, $v$ and $p$. This arrangement is illustrated in \Cref{fig-staggered-grid-01}.
\end{itemize}

To discretely represent the derivatives in \Cref{eqn-navier-2d}, Taylor series expansions are used:

\begin{equation}
  u(x+\Delta x, y+\Delta y) = u(x,y) + \Delta x \frac{\partial u}{\partial x}
                                     + \Delta y \frac{\partial u}{\partial y}
                                     + \mathcal O(\Delta x^2, \Delta y^2, \Delta x \Delta y)
  \label{eqn-dudxdy}
\end{equation}

For Cartesian lattice meshes, this gives:
\begin{equation}
  \begin{array}{rcl}
    \frac{\partial u}{\partial x} &=& \frac{u(x+\Delta x, y) - u(x, y)}{\Delta x} + \mathcal O(\Delta x) \\ [9pt]
    \left.\frac{\partial u}{\partial x} \right|_{i+\frac{1}{2},j}
                                  &\simeq& \frac{u_{i+1,j} - u_{i,j}}{x_{i+1} - x_{i}}
  \end{array}
  \label{eqn-dudx}
\end{equation}

The number of mesh nodes in the discrete meshes is determined by the user and is driven by the
complexity of the geometry being modelled and the solution accuracy required.
As \Cref{eqn-dudx} shows the smaller the mesh spacing, $\Delta x$ the more accurate the Taylor
series approximation. Hence, a dense mesh with a large number of closely spaced nodes gives a more
accurate solution than a coarse mesh with the nodes widely spaced.

\begin{figure}[h]
  \begin{center}
    \includegraphics[width=0.95\textwidth]{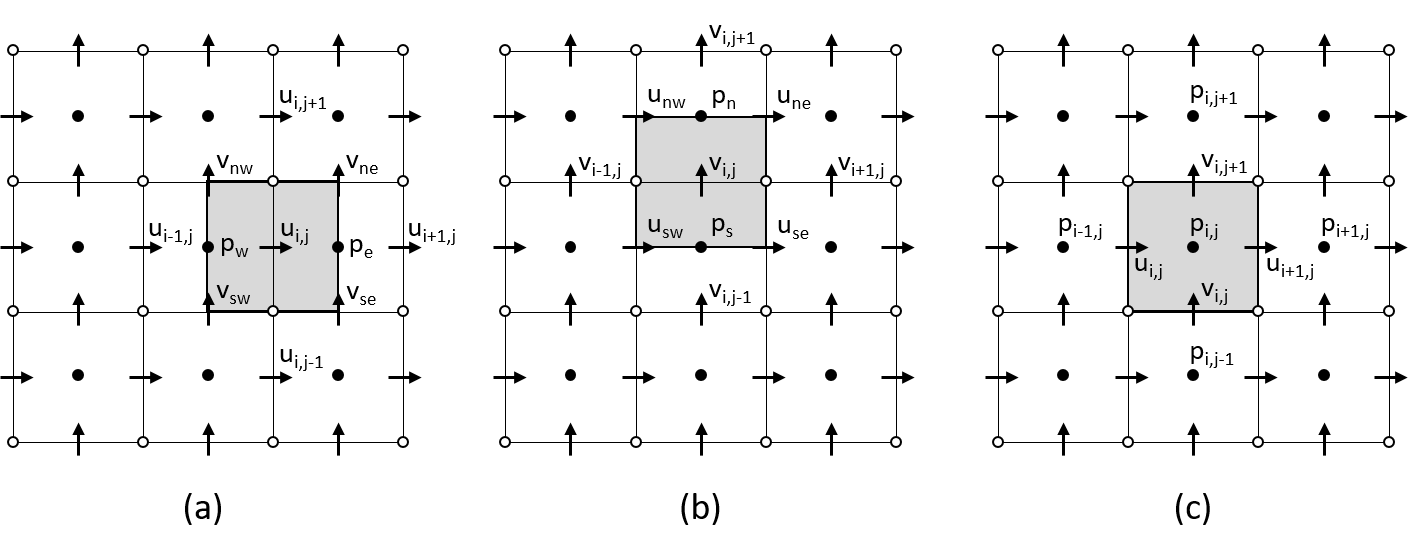}
  \end{center}
  \caption{\small Control volume arrangement for staggered meshes.
           (a) u-velocity control volume, (b) v-velocity control volume, and (c) pressure control volume.}
  \label{fig-staggered-grid-02}
\end{figure}

Repeating the Taylor series expansions for the other derivatives yields a set of discrete
equations in terms of the values of $u, v, p$ at the nodes of their respective 
meshes,  $\mathcal M_u$, $\mathcal M_v$ and $\mathcal M_p$.
To solve the discrete equations the {\it finite volume} method is used.
Details of the methodology are given in Chapter 6 of \cite{versteeg2007introduction}.
The control volumes for each variable are shown in \Cref{fig-staggered-grid-02}.

%
\subsection{Momentum conservation equations}

For the $u$-momentum in \Cref{eqn-navier-2d}, integrating over the control volume
in \Cref{fig-staggered-grid-02} and using the divergence theorem gives:

\begin{equation}
    \int_{e,w} \rho uu dy + \int_{n,s} \rho vu dx =
    -\int_{e,w} p dy +
    \int_{e,w} \mu \frac{\partial u}{\partial x} dy +
    \int_{n,s} \mu \frac{\partial u}{\partial y} dx
  \label{eqn-disc-u-01}
\end{equation}

Where the integrals labelled $e,w$ are over the east and west sides of the face and those
labelled $n,s$ are over the north and west faces. This form is possible due to the Cartesian
nature of the control volume.
Noting that the face normals in the divergence theorem point in opposite directions on opposite
faces, \Cref{eqn-disc-u-01} can be approximated using the discrete mesh values as:

\begin{equation}
  \begin{split}
    \rho u_e u_e \Delta y_e - \rho u_w u_w \Delta y_w +
    \rho v_n u_n \Delta x_n - \rho v_s u_s \Delta x_s = 
    -(p_e \Delta y_e - p_w \Delta y_w) + \\
    \mu \frac{u_{i+1,j} - u_{i,j}}{\Delta x_e} - \mu \frac{u_{i,j} - u_{i-1,j}}{\Delta x_w} +
    \mu \frac{u_{i,j+1} - u_{i,j}}{\Delta y_n} - \mu \frac{u_{i,j} - u_{i,j-1}}{\Delta y_s}
  \end{split}
  \label{eqn-disc-u-02}
\end{equation}

One aspect of using a staggered mesh arrangement is that there isn't a consistent $(i,j)$ indexing
of the $\mathcal M_u$, $\mathcal M_v$ and $\mathcal M_p$ meshes.
In \Cref{eqn-disc-u-02} the $(i,j)$ indices refer to the
$\mathcal M_u$ mesh and the $v$ and $p$ values are labelled according to their cardinal positions
on the $u$-momentum control volume.

The terms such as $u_e u_e$ in \Cref{eqn-disc-u-02} are non-linear terms.
In the SIMPLE algorithm these are treated using a fixed point iteration, also known as a Picard iteration,
where the product $u_e u_e$
is separated  into a {\it convecting} velocity and a {\it convected} velocity.
The {\it convecting} velocity is fixed based on the current solution field and the discrete
equations are used to solve for an updated {\it convected} velocity.
The convention in the SIMPLE method is to label the updated velocity with an asterisk.
As \Cref{sec-app-simple} will show these velocities are interim solutions within on 
overall predictor-corrector algorithm.

Whilst the $\Delta x$ and $\Delta y$ do not vary across a given control volume, the mesh spacing
may be non-uniform meaning, for example, that $\Delta x_e \ne \Delta x_w$.
However, for notational simplicity, $\Delta x$ and $\Delta y$ can be taken as constants.
Hence, applying the fixed point iteration to \Cref{eqn-disc-u-02} gives:

\begin{equation}
  \begin{split}
    \frac{1}{2}\rho u_e \left( u^*_{i,j} + u^*_{i+1,j} \right) \Delta y -
    \frac{1}{2}\rho u_w \left( u^*_{i,j} + u^*_{i-1,j} \right) \Delta y + \\
    \frac{1}{2}\rho v_n \left( u^*_{i,j} + u^*_{i,j+1} \right) \Delta x -
    \frac{1}{2}\rho v_s \left( u^*_{i,j} + u^*_{i,j-1} \right) \Delta x = \\
    -(p^*_e - p^*_w) \Delta y + \\
    \mu \frac{u^*_{i+1,j} - 2u^*_{i,j} + u^*_{i-1,j}}{\Delta x} +
    \mu \frac{u^*_{i,j+1} - 2u^*_{i,j} + u^*_{i,j-1}}{\Delta y}
  \end{split}
  \label{eqn-disc-u-03}
\end{equation}

The {\it convecting} velocities $u_e$, $u_w$, $v_n$ and $v_s$ have not been expanded as this
requires certain stability considerations to be taken into account.
These are described fully in sections 5.6 and onwards of 
\cite{versteeg2007introduction}. For simplicity, this work used the $1^{st}$ order upwinding scheme.

The terms in \Cref{eqn-disc-u-03} can be collected together to give the discrete $u$-momentum
equation:

\begin{equation}
    a^u_P u^*_{i,j} = a^u_E u^*_{i+1,j} + a^u_W u^*_{i-1,j} +
                      a^u_N u^*_{i,j+1} + a^u_S u^*_{i,j-1} -
                     (p^*_e - p^*_w) \Delta y \\
  \label{eqn-disc-u-04}
\end{equation}

The subscripts on the coefficients $a^u$ are capitalised to indicate they correspond to
nodes in the mesh $\mathcal M_u$.
Using $nb$ to denote the cardinal neighbours, \Cref{eqn-disc-u-04} can be simplified to:
\begin{equation}
    a^u_P u^*_{i,j} = \sum_{nb} a^u_{nb} u^*_{nb} - (p^*_e - p^*_w) \Delta y 
  \label{eqn-disc-u-05}
\end{equation}

Following the same steps on mesh $\mathcal M_v$, the discrete $v$-momentum
equation is:

\begin{equation}
    a^v_P v^*_{i,j} = \sum_{nb} a^v_{nb} v^*_{nb} - (p^*_n - p^*_s) \Delta x 
  \label{eqn-disc-v-01}
\end{equation}

\Crefrange{eqn-disc-u-05}{eqn-disc-v-01} illustrate the main reason for using a staggered
mesh which is that the pressure gradients in each equation are computed using the difference
between adjacent pressure values in the $\mathcal M_p$ mesh.
This avoids a pressure-velocity decoupling that can create a spurious chequerboard
pattern in the pressure field \cite{harlow1965numerical}.

%
\subsection{Mass conservation equation}

\Crefrange{eqn-disc-u-05}{eqn-disc-v-01} can solved using a number of iterative schemes
for $u^*$ and $v^*$ with fixed values for $p^*$.
The next step in the SIMPLE algorithm is to use the mass conservation equation to
{\it correct} the velocities and pressure by:

\begin{equation}
  \begin{array}{rcl}
    u &=& u^* + u^{\prime} \\
    v &=& v^* + v^{\prime} \\
    p &=& p^* + p^{\prime}
  \end{array}
  \label{eqn-cont-01}
\end{equation}

The corrections should produce a solution that satisfies the discrete momentum
equations and since the starred variables already do so, substituting \Cref{eqn-cont-01}
into \Crefrange{eqn-disc-u-05}{eqn-disc-v-01} produces:

\begin{equation}
  \begin{array}{rcl}
    a^u_P u^{\prime}_{i,j} &=& \sum_{nb} a^u_{nb} u^{\prime}_{nb} - (p^{\prime}_e - p^{\prime}_w) \Delta y \\[9pt]
    a^v_P v^{\prime}_{i,j} &=& \sum_{nb} a^v_{nb} v^{\prime}_{nb} - (p^{\prime}_n - p^{\prime}_s) \Delta x 
  \end{array}
  \label{eqn-cont-02}
\end{equation}

Note that the updates
are part of of an overall iterative scheme where in the converged solution:
$|u^{\prime}|, |v^{\prime}|, |p^{\prime}| < \epsilon_{tol}$ with $\epsilon_{tol}$
being a small number close to machine precision.
Hence the sums of neighbours in \Cref{eqn-cont-02} can be ignored knowing that in
the converged solution they will be close to zero,
Replacing the the cardinal references for $p$ with the mesh indexing as shown 
\Cref{fig-staggered-grid-02}c yields:
\begin{equation}
  \begin{array}{rcl}
    u_{i,j} &=& u^*_{i,j} - \frac{\Delta y}{a^u_{i,j}} (p^{\prime}_{i,j} - p^{\prime}_{i-1,j}) \\ [9pt]
    v_{i,j} &=& v^*_{i,j} - \frac{\Delta x}{a^v_{i,j}} (p^{\prime}_{i,j} - p^{\prime}_{i,j-1}) 
  \end{array}
  \label{eqn-cont-03}
\end{equation}

Discretising the mass conservation equation using the divergence theorem gives:
\begin{equation}
  \rho (u_{i+1,j} - u_{i,j}) \Delta y + \rho(v_{i,j+1} - v_{i,j}) \Delta x = 0
  \label{eqn-cont-04}
\end{equation}

Substituting $u$ and $v$ from \Cref{eqn-cont-03} into \Cref{eqn-cont-04} gives an
equation of the form
\begin{equation}
  a^p_P p^{\prime}_{i,j} = \sum_{nb} a^p_{nb} p^{\prime}_{nb} -
  \rho \left( (u_{i+1,j}^* - u_{i,j}^*) \Delta y + (v_{i,j+1}^* - v_{i,j}^*) \Delta x \right)
  \label{eqn-pc-01}
\end{equation}

The term involving $u^*$ and $v^*$ on the RHS of \Cref{eqn-pc-01} can be seen to be the deficit
by which the solution of the momentum equations satisfies the continuity equations, usually referred
to as the residual error or just residual.
If the continuity residual is zero, then $p^{\prime}=0$ and both the momentum and mass conservation
equations have been solved.

%
\subsection{The SIMPLE algorithm}
\label{sec-app-simple}
Before summarising the SIMPLE algorithm, the notation is changed to use the Dirac {\it bra-ket} notation.
Let $\ket{u}$ be the vector of $u_{i,j}$ velocities $\forall (i,j) \in \mathcal M_u$,
Similarly for $\ket{v}$, $\ket{p}$ and the $\mathcal M_v$, $\mathcal M_p$ meshes respectively.
The point-wise discretisation equations for $u^*$ and $v^*$ from \Crefrange{eqn-disc-u-05}{eqn-disc-v-01}
and the continuity \Cref{eqn-pc-01}
for every node in their meshes can be written as matrix equations:

\begin{equation}
  \begin{array}{rcl}
    A^u \ket{u^*}        &=& D^x \ket{p^*} \\
    A^v \ket{v^*}        &=& D^y \ket{p^*} \\
    A^p \ket{p^{\prime}} &=& M^x \ket{u^*} + M^y \ket{v^*} \\
  \end{array}
  \label{eqn-simple-01}
\end{equation}

where the entries in matrices $A^u$, $A^v$ and $A^p$ contain linerised terms depending on
$\ket{u}$, $\ket{v}$ and $\ket{p}$.
The matrices $D^x$ and $D^y$ depend only on the mesh coordinates; and
$M^x$ and $M^y$ depend on the mesh coordinates and the constant density.
\Cref{alg-simple} gives the full SIMPLE method.

\begin{algorithm}
  \SetAlgoLined
  \KwResult{Solution $\ket{u}$, $\ket{v}$ and $\ket{p}$ to the 2D mass and momentum equations}
   Set initial values for $\ket{u}$, $\ket{v}$ and $\ket{p}$\;
   converged $\gets 0$\;
   \While{While converged $\neq 0$}{
     set $A^u$ and $A^v$ from $\rho$, $\ket{u}$, $\ket{v}$ and set $\ket{p^*} \gets \ket{p}$\;
     solve\;
     $A^u \ket{u^*} = D^x \ket{p^*}$ on the mesh $\mathcal M_u$\;
     $A^v \ket{v^*} = D^y \ket{p^*}$ on the mesh $\mathcal M_v$\;
     set $A^p$ from $A^u$ and $A^v$\;
     solve\;
     $A^p \ket{p^{\prime}} = M^x \ket{u^*} + M^y \ket{v^*}$ on the mesh $\mathcal M_p$\;
     $\ket{u} \gets \ket{u^*} + \ket{u^{\prime}}$\;
     $\ket{v} \gets \ket{v^*} + \ket{v^{\prime}}$\;
     $\ket{p} \gets \ket{p^*} + \ket{p^{\prime}}$\;
     \If{$\braket{u^{\prime}}, \braket{v^{\prime}}, \braket{p^{\prime}} \le \epsilon^2_{tol}$}{
       converged $\gets 1$\;
     }
   }
  \caption{SIMPLE (Semi-Implicit Method for Pressure Linked Equations)}
  \label{alg-simple}
\end{algorithm}

%
\subsection{Boundary conditions}

There are 2 general classes of boundary conditions used in CFD:
\begin{itemize}
  \item{\bf Dirichlet} - the value of a variable is prescribed at the boundary.
  \item{\bf Neumann} - the derivative of a variable is prescribed at the boundary.
\end{itemize}

These are applied over the set of mesh nodes, $\partial \mathcal M_u$, that form the boundary of $\mathcal M_u$.
In general, boundaries are topologically defined, e.g. inflow, outflow, wall, and there are multiple
nodes on each boundary and multiple boundaries of each type in a CFD calculation.
If there are $n$ such boundaries:

\begin{equation}
  \partial \mathcal M_u = \bigcup_{k=1}^{n} \partial \mathcal M_u^k
  \label{eqn-bc-01}
\end{equation}

Where each boundary set $\partial \mathcal M_u^k$ contains all the mesh nodes that lie
on that boundary and share the same boundary type.
In this work, only Dirichlet boundary conditions are used. 
For example, if node labelled $(i,j)$ is on a solid wall boundary moving with velocity $u_{wall}$:

\begin{equation}
  \begin{array}{rcl}
    u^*_{i+1,j}           &=& u_{wall} \\
    u^{\prime}_{i+1,j}  &=& 0
  \end{array}
  \label{eqn-bc-02}
\end{equation}

In matrix form, the boundary conditions are included as:
\begin{equation}
  \begin{array}{rcl}
    (A^u + B^u) \ket{u^*} &=& D^x \ket{p^*} + \ket{u_b} \\
    (A^v + B^v) \ket{v^*} &=& D^y \ket{p^*} + \ket{v_b} \\
  \end{array}
  \label{eqn-bc-03}
\end{equation}

Where $\ket{u_b}$ and $\ket{v_b}$ have zero amplitudes at the interior nodes
and the specified boundary values at the boundary nodes.

The construction of the staggered mesh means that boundary conditions for the
pressure are not needed. However, since the pressure only enters the
Navier-Stokes equations via its gradients, it is only is only unique up to an
additive constant. To ensure that the solution of the pressure correction
remains bounded, \Cref{eqn-pc-01} is modified at one node in the $\mathcal M_p$
mesh, usually the node (1,1), to be:

\begin{equation}
  a^p_P p^{\prime}_{1,1} = 0 
  \label{eqn-bc-04}
\end{equation}

This is achieved by setting all the neighbour coefficients $a^p_{nb}$ to zero at node (1,1).

%
\section{Generating a linear combination of unitaries for CFD matrices}
\label{sec-app-lcu-method}

As shown in \Cref{sec-app-cfd-alg}
CFD algorithms create inner linearised systems that solve an update equation of the
form $A \ket{\delta x} =\ket{\delta b}$, where $A$ and $\ket{\delta b}$ depend on $\ket{x}$.
After solving the linear system: $\ket{x} \gets \ket{x} + \ket{\delta x}$,
$A$ and $\ket{\delta b}$ are updated using the new value of $\ket{x}$ and the process is 
repeated until $\bra{\delta x}\ket{\delta x}$ and/or $\bra{\delta b}\ket{\delta b}$ are
below a user specified convergence tolerance.
The characteristic feature of this process is that the entries in $A$ change each time
$\ket{x}$ is updated but its sparsity pattern does not.

A new method for rapidly recalculating the coefficients in a LCU is discussed in this Appendix.
Although the finite volume discretisation scheme used in CFD creates symmetric entries, the boundary
conditions add non-symmetric terms.
Hence, at each iteration, the linear system to be solved is:
\begin{equation}
  \begin{pmatrix}
    0           & A \\
    A^{\dagger} & 0
  \end{pmatrix}
  \begin{pmatrix}
    0 \\
    \delta x \\
  \end{pmatrix}
  =
  \begin{pmatrix}
    \delta b \\
    0
  \end{pmatrix}
  \label{eqn-app-Hx-b}
\end{equation}

Which requires a unitary decomposition of the form:
\begin{equation}
  H = \sum_{i=1}^{M} \alpha_i U_i
  \label{eqn-app-lcu01}
\end{equation}

Since, the focus of this study is CFD matrices only real valued matrices are considered.
We also assume that there are few, if any, repeated entries in a CFD matrix other than
pairwise symmetric entries.

%
\subsection{Entry-wise decomposition}
\label{subsec-app-entryLCU}
A straightforward approach is to use each pair of symmetric entries in $H$ as the basis for
a pair of 1-sparse unitaries. If $h_{i,j} = h_{j,i}$ are the symmetric pair, initialise two
unitaries with unit entries at the same locations. A simple marching algorithm can then
add $1$ to each remaining row and column of one of the unitaries and $-1$ to the second.
The coefficient for each unitary is $\frac{1}{2}h_{i,j}$.
This is illustrated in \Cref{eqn-app-unit00} 

\begin{equation}
  \begin{pmatrix}
    0   & 0   & 0   & 0 \\
    0   & 0   & 0.4 & 0 \\
    0   & 0.4 & 0   & 0 \\
    0   & 0   & 0   & 0 \\
  \end{pmatrix}
  =
  0.2
  \begin{pmatrix}
    1   & 0   & 0   & 0 \\
    0   & 0   & 1   & 0 \\
    0   & 1   & 0   & 0 \\
    0   & 0   & 0   & 1 \\
  \end{pmatrix}
  +0.2
  \begin{pmatrix}
   -1   & 0   & 0   & 0 \\
    0   & 0   & 1   & 0 \\
    0   & 1   & 0   & 0 \\
    0   & 0   & 0   &-1 \\
  \end{pmatrix}
  \label{eqn-app-unit00}
\end{equation}

This approach is the most efficient in terms of re-evaluating the coefficients
of the unitaries as there is a direct 1-1 correspondence between the coefficients and
the entries in the CFD matrix.
However, there is some arbitrariness in the creation of unitaries and not all are guaranteed to 
produce efficient circuit implementations.
Similar methods are more effective on matrices with many repeated entries.
\cite{vazquez2018quantum} used a 1-dimensional colouring scheme to decompose a
tri-diagonal Toeplitz matrix into three 1-sparse matrices with non-integer coefficients.
\cite{zhou2017efficient} decomposed dense circulant matrices into a sum of 1-sparse
matrices each of which corresponded to one of the repeated coefficients in the 
circulant matrix. Circuits were written in terms of the unitaries and not further
decomposed.
\cite{ahokas2004improved} provided circuits for a number of 1-sparse Hamiltonians
including those with real valued entries. However, these circuits required $n$ ancilla
qubits in addition the $n$ qubits on which the Hamiltonian operated.

For circuit optimisation, it
is preferable to decompose the Hamiltonian into unitaries that are products of the Pauli
operators such as in the Jordan-Wigner transformation used to relate Pauli
strings with Fermionic operators in quantum chemistry \cite{nielsen2005fermionic}.
However, CFD matrices do not have a Fermionic basis and an alternative method of
creating Pauli strings is needed. 
Note that the matrix in \Cref{eqn-app-unit00} can alternatively be decomposed into:
$0.2 X \otimes X + 0.2 Y \otimes Y$.

%
\subsection{Decomposition using trace orthogonality of Pauli matrices}
\label{subsec-app-pauli-trace}
If each unitary in \Cref{eqn-app-lcu01} is a product of Pauli matrices, each can be
written as:

\begin{equation}
  U_i = \bigotimes_{j=1}^{n} P_{i,j}
  \label{eqn-pprod01}
\end{equation}

where $n = log_2(N)$ is the number of qubits needed for the unitary register,
$N$ is the rank of $H$ assumed to be a power of 2 and $P_{i,j} \in \{I, X, Y, Z\}$
is one of the Pauli matrices (extended to include the identity matrix):

\begin{equation}
  \begin{array}{rcl}
    I &=& \begin{pmatrix}
             1 &  0 \\
             0 &  1 
          \end{pmatrix} \\[10pt]
    X &=& \begin{pmatrix}
             0 &  1 \\
             1 &  0 
          \end{pmatrix} \\[10pt]
    Y &=& \begin{pmatrix}
             0 & -i \\
             i &  0 
          \end{pmatrix} \\[10pt]
    Z &=& \begin{pmatrix} 
             1 &  0 \\
             0 & -1 
          \end{pmatrix}
  \end{array}
  \label{eqn-pauli01}
\end{equation}

The number of possible combinations of Pauli matrices is $4^n$, or $N^2$,
i.e. the number of entries in $H$ if it were fully populated.

Trace orthogonality of Pauli matrices arises from the condition:

\begin{equation}
  \frac{1}{2}Tr(P_iP_j) = \delta_{ij}
  \label{eqn-pauli-trace01}
\end{equation}

Since $Tr(A\otimes B) = Tr(A) Tr(B)$ and
$(A\otimes B)(C\otimes D) = (AC) \otimes (BD)$
trace orthogonality extends to products of Pauli matrices:

\begin{equation}
  U_i U_j = \left(\bigotimes_{l=1}^{n} P_{i,l} \right) 
            \left(\bigotimes_{m=1}^{n} P_{j,m} \right)
          = \bigotimes_{l=1}^{n} P_{i,l} P_{j,l} 
  \label{eqn-pauli-trace02}
\end{equation}

Giving,
\begin{equation}
  Tr(U_i U_j) = \prod_{l=1}^{n} Tr(P_{i,l} P_{j,l}) = 2^n \prod_{l=1}^{n} \delta_{i_lj_l}
  \label{eqn-pauli-trace03}
\end{equation}

Hence each coefficient in a unitary expansion based on Pauli strings can be found using:

\begin{equation}
 \alpha_i = \frac{1}{2^n} Tr(U_i H)  
  \label{eqn-pauli-trace04}
\end{equation}

Calculating which unitaries are in the decomposition notionally requires all $N^2$ 
Pauli products to be checked. Care is needed on the initial set-up to ensure starting values
which happen to be equal do not cause unitaries to be omitted that would have non-zero
coefficients in later iterations.

Re-evaluating the LCU coefficients is a simple matter of reapplying \Cref{eqn-pauli-trace04}
using the stored unitaries. Using 64-bit integers and double precision reals this requires
$3MN \times 8$ bytes of storage. This is independent of the sparse storage format as they all have
to store the row, column and value of each 1-sparse entry.
For this study, storing the unitaries is not an issue and provides a like-with-like
comparison with the method that has been developed and will be described next.

%
\subsection{Decomposition using Hadamard based orthogonality of Pauli matrices}
\label{subsec-app-pauli-hadamard}
In this work, an alternative form of orthogonality has been developed based on the
grand sum of Hadamard products \cite{lapworth2022patent}.
The Hadamard product, denoted by $\circ$, is the element-wise product of 2 matrices of
the same shape. For example:

\[
  \begin{pmatrix}
    a &  b \\
    c &  d 
  \end{pmatrix}
  \circ
  \begin{pmatrix}
    e &  f \\
    g &  h 
  \end{pmatrix}
  = 
  \begin{pmatrix}
    ae &  bf \\
    cg &  dh 
  \end{pmatrix}
\]

More generally, if $A$ and $B$ have the same shape:

\begin{equation}
  (A \circ B)_{ij} = A_{ij} B_{ij}
  \label{eqn-hadam01}
\end{equation}

From \Cref{eqn-pauli01} and \Cref{eqn-hadam01}, the Hadamard
products of the Pauli matrices can be calculated as shown in \Cref{tab-hadam01}.

\begin{table}[ht]
  \centering
  \begin{tabular}{c|c c c c}
    $\circ$ & I & X & Y & Z \\
    \hline
       I    & I & 0 & 0 & Z \\
       X    & 0 & X & Y & 0 \\
       Y    & 0 & Y &-X & 0 \\
       Z    & Z & 0 & 0 & I \\
  \end{tabular}
  \caption{Hadamard products of the Pauli matrices. Note that 0 is the 2x2 null matrix
           with all zero entities.}
  \label{tab-hadam01}
\end{table}

Define $G$ as the {\it grand sum} of all the elements in a matrix:
\begin{equation}
  G(A) = \sum_{i,j} A_{i,j}
  \label{eqn-gsum01}
\end{equation}

And $G_H$ the grand sum of a Hadamard product:

\begin{equation}
  G_H(A,B) = G(A \circ B) = \sum_{i,j} A_{i,j}B_{ij}
  \label{eqn-gsum02}
\end{equation}

Applying $G_H$ to the Hadamard products in \Cref{tab-hadam01} gives:

\begin{table}[ht]
  \centering
  \begin{tabular}{c|c c c c}
    $G_H$   & I & X & Y & Z \\
    \hline
       I    & 2 & 0 & 0 & 0 \\
       X    & 0 & 2 & 0 & 0 \\
       Y    & 0 & 0 &-2 & 0 \\
       Z    & 0 & 0 & 0 & 2 \\
  \end{tabular}
  \caption{Grand sum of Hadamard products of the Pauli matrices.}
  \label{tab-gsum01}
\end{table}

As \Cref{tab-gsum01} shows, $G_H$ provides an orthogonal operator for
Hadamard products of Pauli matrices. As will be shown later, under the conditions
of a typical CFD matrix, this can be extended to provide orthonormality.

From their definitions, Hadamard and Kronecker products are related by:

\begin{equation}
  (A \otimes B) \circ (C \otimes D) = (A \circ C) \otimes (B \circ D)
  \label{eqn-hadk01}
\end{equation}

By construction:

\begin{equation}
  G(A \otimes B) = G(A)G(B)
  \label{eqn-hadk02}
\end{equation}

Hence:
\begin{equation}
  G_H(A \otimes B, C \otimes D) = G_H(A,C)G_H(B,D)
  \label{eqn-hadk03}
\end{equation}

%
\subsubsection{Sparsity considerations}

Let $S_H$ denote the sparsity pattern of the symmetrised CFD matrix and 
$S_U$ denote the sparsity pattern of the sum of unitaries in \Cref{eqn-app-lcu01}.
Since, the objective of the methodology is to re-use the same set of unitaries for 
matrices with different entries during the CFD solution process, the decomposition must ensure
that $S_H \cap S_U = S_H$.
In general, the relative complement $S_U - S_H$ is not an empty set and contains some of the zero entries 
in $H$ that are not stored in its sparse matrix representation. 
The solution of \Cref{eqn-app-lcu01} must ensure these entries are zero.
To do this, the sparse matrix representation of $H$ is in-filled with the zero entries to create an expanded
matrix $\hat{H}$ such that $S_U - S_{\hat{H}} = \{\}$. 
\Cref{eqn-app-lcu01} can be restated as:
\begin{equation}
  \hat{H} = \sum_{i=1}^{\hat{M}} \alpha_i U_i
  \label{eqn-lcu02}
\end{equation}

Mathematically, \Cref{eqn-app-lcu01} and \Cref{eqn-lcu02} are equivalent, the distinction between
$H$ and $\hat{H}$ is purely algorithmic due to their different sparsity patterns.
Typically, $\hat{M} \ge M$, because the algorithm computes the unitary decomposition for all
possible non-zero entries rather than a specific set of values.
For any given sparse matrix, some or many of the $\alpha_i$ values in \Cref{eqn-lcu02} may be 
zero.

A second sparsity consideration is that I and Z, and X and Y have the same sparsity patterns.
It is easy to see that in the set of all $n$ qubit tensor products of Pauli matrices, there
are subsets of size $N=2^n$ that have the same sparsity pattern.
Since all products of Pauli matrices are {\it 1-sparse}, the number of non-zero entries in
each product is also $N$.
The intersection of any 2 subsets is the empty set, i.e. the subsets have non-overlapping
sparsity patterns that fill a square matrix of rank $N$.
This can be seen by the fact that such a matrix has $N^2$ entries which is the same as the
number of subsets times the number of non-zero entries in the shared sparsity pattern of
each set. If the sparsity pattern of two subsets overlapped, there would be at least one 
position in the matrix not included in any of the sparsity subsets.

%
\subsubsection{Solving for the unitary coefficients}
Rather than solve for the complete set of unitaries, it is more efficient to solve
for each subset. Since all the elements of a subset share a common sparsity pattern, 
the subsets are called clusters. Each cluster contains M unitaries each with N non-zero
entries. Applying the {\it vec} operator to the non-zero entries in each unitary gives:

\begin{equation}
  \begin{pmatrix}
    \hat{H}_1 \\
    \hat{H}_2 \\
    \vdots \\
    \hat{H}_N 
  \end{pmatrix}
  = 
  \begin{pmatrix}
    U_{1,1} & U_{2,1} & \ldots & U_{M,1} \\
    U_{1,2} & U_{2,2} & \ldots & U_{M,2} \\
    \vdots  & \vdots  & \ddots & \vdots  \\
    U_{1,N} & U_{2,N} & \ldots & U_{M,N} 
  \end{pmatrix}
  \begin{pmatrix}
    \alpha_1 \\
    \alpha_2 \\
    \vdots \\
    \alpha_M 
  \end{pmatrix}
\label{eqn-alpha01}
\end{equation}

\Cref{eqn-alpha01} uses a local indexing following the sparse storage method of the
matrices, e.g. compressed row, compressed column, etc.
In this equation $U_{i,j}$ denotes the $j^{th}$ entry in the $i^{th}$ unitary.
It appears that all that has happened is that an $N \times M$ full matrix inversion is
needed to solve for the coefficients. However, denoting the matrix of unitaries 
by $U$ gives:

\begin{equation}
  U^T U =
  \begin{pmatrix}
    U_{1,1} & U_{1,2} & \ldots & U_{1,N} \\
    U_{2,1} & U_{2,2} & \ldots & U_{2,N} \\
    \vdots  & \vdots  & \ddots & \vdots    \\
    U_{M,1} & U_{M,2} & \ldots & U_{M,N} 
  \end{pmatrix}
  \begin{pmatrix}
    U_{1,1} & U_{2,1} & \ldots & U_{M,1} \\
    U_{1,2} & U_{2,2} & \ldots & U_{M,2} \\
    \vdots  & \vdots  & \ddots & \vdots    \\
    U_{1,N} & U_{2,N} & \ldots & U_{M,N} 
  \end{pmatrix}
\label{eqn-alpha02}
\end{equation}

Since all the unitaries have the same sparsity pattern and the omitted entries are
all zero, the product of the $i^{th}$ row  of $U^T$  and the $j^{th}$ column of $U$
is the grand sum of a Hadamard product:

\[
(U^T U)_{ij} = G(U_i \circ U_j) = G_H(U_i , U_j)
\]

Using the definition of $U$ from \Cref{eqn-pprod01} and the identities from
\Cref{eqn-hadk01} and \Cref{eqn-hadk02} gives:

\begin{equation}
  \begin{split}
   (U^T U)_{ij} & =  G \left(\bigotimes_{k=1}^{n} P_{i,k} \circ \bigotimes_{k=1}^{n} P_{j,k} \right) \\
                & =  G \left(\bigotimes_{k=1}^{n} (P_{i,k} \circ P_{j,k}) \right) \\
                & = \prod_{k=1}^{n} G_H \left(P_{i,k} , P_{j,k} \right)
  \end{split}
  \label{eqn-uut01}
\end{equation}

Considering $(U^T U)_{ii}$ and using the values from \Cref{tab-gsum01} gives:
\begin{equation}
  (U^T U)_{ii} = \prod_{k=1}^{n} (\pm 2) = \pm 2^n 
  \label{eqn-uut02}
\end{equation}

However, it is only $Y \circ Y$ that results in $-2$.
If the methodology is restricted to only real-valued CFD matrices, then each
$U_i$ must contain an even number of $Y$ matrices and the final product in
\Cref{eqn-uut01} must contain and even number of $Y \circ Y$ products. Hence,
\begin{equation}
  (U^T U)_{ii} = 2^n, \; \forall i
  \label{eqn-uut03}
\end{equation}

If $i \ne j$, then the since unitaries in the cluster are unique, there must be
at least one 
Hadamard product $P_{i,k} \circ P_{j,k}$ for which two different Pauli matrices are
being multiplied. From \Cref{tab-gsum01}, there is at least one
$G_H \left(P_{i,k} , P_{j,k} \right) = 0$. Hence,
\begin{equation}
  (U^T U)_{ij} = 0, \; \forall i \ne j
  \label{eqn-uut04}
\end{equation}

Combining \Cref{eqn-uut03} and \Cref{eqn-uut04} gives:
\begin{equation}
  U^T U = 2^n I
  \label{eqn-uut05}
\end{equation}

And the coefficients for this cluster can be directly computed from:
\begin{equation}
  \begin{pmatrix}
    \alpha_1 \\
    \alpha_2 \\
    \vdots \\
    \alpha_M 
  \end{pmatrix}
  = 
  \frac{1}{2^n}
  \begin{pmatrix}
    U_{1,1} & U_{1,2} & \ldots & U_{1,N} \\
    U_{2,1} & U_{2,2} & \ldots & U_{2,N} \\
    \vdots  & \vdots  & \ddots & \vdots    \\
    U_{M,1} & U_{M,2} & \ldots & U_{M,N} 
  \end{pmatrix}
  \begin{pmatrix}
    \hat{H}_1 \\
    \hat{H}_2 \\
    \vdots \\
    \hat{H}_N 
  \end{pmatrix}
  \label{eqn-alpha03}
\end{equation}

Repeating this process for all clusters gives the complete unitary decomposition for $H$.
Storing $U^T$ for each cluster allows the unitary decomposition to be quickly recomputed
each time the outer CFD iteration updates the matrix system to be solved.

Note, that in the above derivation $M$ and $N$ do not have to be equal and for 
symmetrised matrices following \Cref{eqn-A2H}, the algorithm can be applied to
the upper right block of $H$ for which $M = \frac{N}{2}$.

An important computational aspect of this result is that the entries $U_{ij}$
all have values of $\pm 1$. 
By letting $0 \mapsto -1$, a single bit, $b$, can be used to store the each entry
such that the its value is $-1 + 2b$.
In the current implementation, a single byte is used, i.e. {\it int8\_t} in C.

Comparing the storage requirements with the trace orthogonality approach, this method effectively
stores all the non-zero entries in the LCU decomposition but not as a set of sparse 
matrices. The total number of entries stored is $MN$ bytes, using the 1 byte for each
entry. Since each cluster shares a sparsity pattern, the sparse matrix indexing needs
only be stored once for each cluster, requiring $2n_cN \times 8$ bytes of storage.
Since $n_c \ll M$ this is a significant saving in storage relative to the
trace orthogonality method.

%
\bibliographystyle{ieeetr}
\bibliography{references}

\end{document}